\documentclass{aa}  
\usepackage{graphicx}
\usepackage{txfonts}
\usepackage[svgnames,
            dvipsnames]{xcolor}
\usepackage[colorlinks=true,
            citecolor=blue,   
            linkcolor=blue,   
            urlcolor=blue]{hyperref}
\usepackage{comment}
\usepackage{xspace}
\usepackage{multirow}
\usepackage{lipsum}
\usepackage{cancel}
\usepackage[normalem]{ulem}
\usepackage[switch]{lineno}

\graphicspath{{Figures/}}

\newcommand{\scipiclite}{\texttt{SciPIC-lite}\xspace}
\newcommand{\Nbody}{$N$-body\xspace}
\newcommand{\hMpc}{h^{-1} \ \mathrm{Mpc}}

\newcommand{\de}{\mathrm{d}}
\newcommand{\der}[2]{\frac{\de#1}{\de#2}}
\newcommand{\wprp}{w_\mathrm{p}}
\newcommand{\xiell}{\xi_\ell}
\newcommand{\deltalogMh}{\Delta_{\log M}}
\newcommand{\logMmin}{\log M_\mathrm{min}~}

\newcommand{\logMone}{\log M_1~}
\newcommand{\Mone}{M_1}
\newcommand{\logMcut}{\log M_\mathrm{cut}~}
\newcommand{\Mcut}{M_\mathrm{cut}}
\newcommand{\sigmalogM}{\sigma_{\log M}~}
\newcommand{\fmax}{f_\mathrm{max}~}
\newcommand{\Mh}{M_\mathrm{h}}
\newcommand{\logMh}{\log M_\mathrm{h}~}
\newcommand{\Ncen}{N_\mathrm{cen}}
\newcommand{\Nsat}{N_\mathrm{sat}}

\newcommand{\mean}[1]{\langle#1\rangle}
\newcommand{\ind}[1]{\mathrm{#1}}
\newcommand{\redchisq}{\ensuremath{{\chi^2_\mathrm{red}}}}

\begin{document} 

\title{\scipiclite: A flexible Halo Occupation Distribution pipeline aimed at clustering observables}



\author{Gabriele Parimbelli\inst{1,2}\fnmsep\thanks{corresponding author}
        \and
        Martin Crocce \inst{1,2}
        \and
        Giosuè Gambardella\inst{1,2}
        \and
        Alexander Eggemeier\inst{3,4}
        \and
        Pablo Fosalba\inst{1,2}
        \and
        Elizabeth J. Gonzalez\inst{5,6}
        \and
        Sujatha Ramakrishnan\inst{1}
        \and
        Isaac Tutusaus\inst{1,2,7}
        \and
        Zahra Baghkhani\inst{1,2}
        \and
        Jorge Carretero\inst{5,6}
        \and
        Francisco Castander\inst{1,2}
        \and
        Andrea Pezzotta\inst{8}
        \and
        Pau Tallada-Cresp\'{i}\inst{5,6}
        }

\institute{
    Institute of Space Sciences (ICE, CSIC), Campus UAB, Carrer de Can Magrans, s/n, 08193 Barcelona, Spain
     \and
    Institut d'Estudis Espacials de Catalunya (IEEC), Edifici RDIT, Campus UPC, 08860 Castelldefels (Barcelona), Spain
     \and
    Universität Bonn, Argelander-Institut für Astronomie, Auf dem Hügel 71, 53121 Bonn, Germany
     \and
    Cluster of Excellence DYNAVERSE
    \and
    Port d’Informació Científica (PIC), Campus UAB, C. Albareda s/n, 08193 Bellaterra (Barcelona), Spain
    \and
    Centro de Investigaciones Energéticas, Medioambientales y Tecnológicas (CIEMAT), Avenida Complutense 40, 28040 Madrid, Spain
    \and
    Univ Toulouse, CNES, CNRS, IRAP, 14 Av.~Edouard Belin, 31400 Toulouse, France
    \and
    INAF - Osservatorio Astronomico di Brera, via Brera 28, 20121 Milan, Italy
\\
\newline
\email{parimbelli@ice.csic.es}
}
\date{Received XXX; accepted YYY}

\abstract
    {The Halo Occupation Distribution (HOD) framework provides a flexible approach to modelling the galaxy--halo connection in cosmological simulations, encoding the statistical relationship between galaxies and their host dark matter haloes.
    We present \scipiclite, a new public Python library for fast, flexible, and user-friendly HOD modelling, designed to populate dark matter haloes extracted from cosmological simulations with galaxies and to provide a comprehensive suite of tools for their subsequent analysis.
    The library is built around a modular and extensible codebase that incorporates the full breadth of the HOD framework, from the core occupation functions to a wide range of physically motivated extensions, making it suitable for a broad spectrum of galaxy clustering applications. We validate and characterise the performance of \scipiclite by developing a pipeline to fit the two-point correlation functions of H$\alpha$-selected galaxies from the Flagship~2 simulation across multiple redshift bins. To this end, we populate AbacusSummit dark matter haloes with galaxies according to a variety of HOD parametrisations. We assess the robustness of the fits with respect to the choice of HOD parametrisation and the inclusion of physically motivated extensions, including assembly bias, velocity bias, conformity, and halo profile rescaling. The best-fit models are found to be remarkably robust across different HOD parametrisations and fitting configurations, with reduced $\chi^2$ values clustering tightly around a common value regardless of the model adopted. We exploit these results to generate new AbacusSummit-based galaxy mocks that accurately reproduce the clustering properties of the Flagship~2 H$\alpha$ galaxy population.}


\maketitle

\section{Introduction}
\label{sec:introduction}

The construction of realistic galaxy mock catalogues from high-resolution cosmological simulations plays a central role in the study of galaxy clustering and the large-scale structure of the Universe. 
Such mocks are indispensable for testing the range of scales over which theoretical models remain valid \citep{DESI_DR1-scale_cuts,DESI_DR1_scale_cuts_compression,PT_challenge,PT_challenge_shape_fit,Euclid_Pk_modelling,Euclid_Pk_Bk_modelling,Euclid_23PCF_modelling}, and for gaining deeper insight into the physical processes that govern galaxy formation and evolution \citep{Baryons_on_sham-Contreras+26,ExtHOD-Jimenez+19,Assembly_bias_Illustris-Hadzhiyska+21,Assembly_bias_Illustris-Yang+26}.
In addition, they provide a controlled framework for quantifying systematic and modelling uncertainties \citep{Euclid_systematics-Monaco+26}, validating analysis pipelines, and generating reliable covariance matrices for large-scale structure observables \citep{Euclid_Large_Mocks,cov_2pcf,cov_pk}.

A key ingredient of this process is the so-called \textit{galaxy-halo connection}, i.e. the relationship between the properties of dark matter haloes and the galaxies that they host.
There exist several approaches to studying the galaxy-halo connection \citep[see e.g.][]{Galaxy-halo_connection}; the simplest of these, referred to as Halo Occupation Distribution (HOD) framework, consists of placing galaxies directly into dark matter haloes following empirical relations \citep[e.g.][]{HOD_Peacock,BerlindWeinberg2002_HOD,BerlindWeinberg2003_HODPhysics,Zheng2005_TheoreticalHOD}.
In its standard form, the HOD framework populates each halo with a number of galaxies drawn from an occupation function that depends solely on halo mass.
Extensions of the basic model can reproduce effects predicted by more detailed approaches, such as sub-halo abundance matching \citep[e.g.][]{SHAM-Conroy+06,Contreras2022_SHAMe}, semi-analytic models \citep[SAM; e.g.][]{Croton2016_SAGE,UniverseMachine} and hydrodynamical simulations.
These include assembly bias, i.e. a dependence of galaxy occupation on halo properties beyond mass, such as formation history, concentration or environment \citep[e.g.][]{Assembly_bias_hydro-Chaves-Montero+16}.
Another example is galactic conformity, i.e. correlated properties between central and satellite galaxies \citep[e.g.][]{Conformity_Illustris-Yuan+25,Assembly_bias_hydro-occupation-Hadzhiyska+23a}.
A further one is velocity bias, i.e. a difference between galaxy velocities and those of the underlying dark matter \citep[e.g.][]{velocity_bias_SDSS-Guo+15}.

In the last two decades, the HOD framework has been widely used to characterise the galaxy-halo connection observed by different surveys, from 2dF \citep{Magliocchetti2003} and SDSS \citep{Zehavi-SDSS+11} to VIPERS \citep{VIPERS} and DESI \citep{DESI_ELG_HOD-Rocher+23,DESI_LRG_HOD-Yuan+23}.
Accurately modelling the HOD of a survey population is crucial for the scientific exploitation of its data: a detailed understanding of the galaxy sample is required to build realistic mock catalogues, which are essential to estimate the statistical and systematic error budgets.
Moreover, it can provide informative priors on large-scale models, which can tighten the resulting cosmological constraints \citep{HOD_informed_priors-Zhang+25}.
The characterisation is usually performed by populating dark matter haloes from high-resolution cosmological simulations and fitting the resulting summary statistics, typically two-point correlation functions (2PCF), to the observed data.
Each evaluation of the model is computationally expensive, so that fast methods to populate haloes are required, within pipelines flexible enough to accommodate a wide range of models.

In this work, we present \scipiclite, a newly developed Python library designed for the fast and flexible modelling of galaxy populations through the HOD framework, with a particular focus on galaxy clustering applications.
\scipiclite is inspired by \textsc{SciPIC} \citep{Scipic-Carretero+17,Scipic-Gonzalez+26}, and inherits its core philosophy while offering a significantly more flexible implementation.
\textsc{SciPIC} was the code employed to generate the Flagship~2 galaxy mock \citep{Flagship-Castander+24}, designed to provide a realistic representation of the galaxy population expected from the \textit{Euclid} survey \citep{Euclid}.
\scipiclite supports a wide variety of HOD parametrisations for both central and satellite galaxies, making it suitable in principle for any galaxy population.
It also natively includes several physically motivated extensions of the standard HOD paradigm.
Being primarily designed for galaxy clustering applications, \scipiclite implements a comprehensive set of functions to measure and model the key statistical quantities involved in such analyses, including halo and galaxy mass functions, 2PCFs and power spectra.
The library is designed with extensibility in mind: support for a broad range of galaxy populations is built-in, and new user-defined HOD functions can be integrated into the existing pipeline, making \scipiclite a versatile tool for the wider galaxy clustering community.
\scipiclite will be made publicly available upon publication of this work\footnote{\url{https://github.com/GabrieleParimbelli/scipiclite}}.
The code is accompanied by worked examples and a suite of notebooks illustrating how to generate simple HOD-based galaxy mocks and set up minimisation pipelines.

A widely used alternative is \textsc{AbacusHOD} \citep{AbacusHOD-Yuan+21}, a highly optimised HOD framework developed for the AbacusSummit simulations \citep{Abacus-Maksimova+21}.
While efficient, it is mostly tailored to the Abacus code data products, which limits its portability to other \Nbody simulations.
Moreover, implementing new occupation functions requires modifying the source code, and, at the time of writing, light-cone outputs are not supported.
In contrast, \scipiclite is designed to be simulation-agnostic: it operates on generic halo catalogues, allows user-defined HOD shapes and satellite profiles without modifying the code, and supports both snapshots and light cones, making it a flexible and portable tool for a broad range of applications.

We validate and characterise the performance of \scipiclite in the context of the \textit{Euclid} spectroscopic survey.
\textit{Euclid} primarily targets H$\alpha$ emitters, a subset of Emission Line Galaxies (ELGs) that are among the most abundant and spectroscopically accessible galaxy populations at $z\sim1$.
We focus on this population, using a sample of H$\alpha$ emitters drawn from the Flagship~2 simulation.
We measure the 2PCFs of this galaxy mock in the redshift range $z=[0.85,1.80]$ and use them as our target data.
We then populate dark matter haloes from the AbacusSummit simulation suite \citep{Abacus-Maksimova+21} according to the HOD models under consideration, in order to fit the resulting 2PCFs to the target data.
We exploit the flexibility of the code to systematically explore a broad range of HOD parametrisations, including assembly bias, velocity bias, conformity, and rescaling of halo density profiles.

This paper is organised as follows.
Sect.~\ref{sec:simulations} describes the \Nbody simulations employed in this work.
Sect.~\ref{sec:scipiclite} presents the \scipiclite code, including its main functionalities, clustering tools, and fitting pipeline.
Sect.~\ref{sec:dataset} introduces the dataset, together with the observables and covariance matrices used in the fits.
In Sect.~\ref{sec:results}, we discuss the results, comparing the performance of the different HOD parametrisations and extensions across all redshift bins.
Sect.~\ref{sec:mocks} presents new AbacusSummit-based galaxy mocks built from the small-scale best-fit HOD parameters, and tests whether they also reproduce the large-scale clustering of the Flagship~2 H$\alpha$ population.
Finally, Sect.~\ref{sec:conclusions} summarises our conclusions.

\section{$N$-body simulations}
\label{sec:simulations}

We briefly describe the two simulation suites used in this work.
For more information, we refer the reader to the relevant literature cited in this manuscript.

\subsection{Abacus}
\label{subsec:simulations.abacus}

The AbacusSummit \Nbody simulation suite \citep[][hereafter Abacus]{Abacus-Maksimova+21} is an extensive set of more than 100 simulations, exploring an eight-dimensional parameter space in the $w_0w_a$CDM model.
All the simulations evolve $6~192^3$ particles within a comoving box of side length 2~000 $h^{-1} \ \mathrm{Mpc}$.
Dark matter haloes are identified through a novel, purpose-built halo finder, \texttt{CompaSO} \citep{Compaso+22}.
This tool combines a phase-space-based particle grouping stage with a subsequent spherical overdensity procedure to robustly identify haloes and determine their properties.

Notably, 25 of these simulations share an identical cosmology (dubbed \texttt{c000}), with parameters consistent with the Planck 2018 results \citep{Planck2018}, listed in Table~\ref{tab:cosmo_sims}.
In this work we employ the first of these realisations as our fiducial halo catalogue, which we will populate using our \scipiclite pipeline during the fitting procedure.
We use halo catalogues at three snapshots corresponding to $z = 0.950$, $1.250$, and $1.625$\footnote{According to the Abacus documentation \url{https://abacussummit.readthedocs.io/en/latest/}, these snapshots correspond to ``secondary redshifts'', i.e. outputs for which the time step was not adjusted to match the exact target scale factor. As a result, the effective redshift differs slightly from the nominal value. However, we find that these deviations are at most of order 0.2\% in terms of the linear growth factor, a negligible difference for our purposes.}.
These redshifts are selected to bracket the range in which \textit{Euclid} is expected to observe H$\alpha$ galaxies.

\subsection{Flagship 2}
\label{subsec:simulations.flagship2}

The Flagship 2 \Nbody simulation was run as the benchmark numerical experiment designed to support the \textit{Euclid} survey \citep{Euclid}.
The simulation was performed using the \texttt{PKDGRAV3} \Nbody code \citep{PKDGRAV} and it follows the gravitational evolution of 16~000$^3$ dark matter particles within a cubic volume of side length 3~600 $h^{-1}$ Mpc.
The assumed cosmology is $\Lambda$CDM, characterised by the parameters listed in Table \ref{tab:cosmo_sims}.
We note that the cosmologies of Flagship~2 and Abacus differ slightly; as shown below, these differences are negligible for our purposes.
Massive neutrinos are incorporated as a smooth component, with a particle–mesh force contribution computed at each time step \citep{Fidler+19}.
This setup yields a mass resolution of $10^9 \ M_\odot/h$.
The gravitational softening length is  $4.5\, h^{-1}$ kpc.
A particle light cone is generated on the fly, extending to 
$z=3$ and as wide as one octant of the sky.
Within this light cone, dark matter haloes are identified down to a minimum mass of $10^{10} \ M_\odot/h$, using the \texttt{ROCKSTAR} phase-space halo finder \citep{ROCKSTAR}.

\begin{table}
\caption{Cosmological parameters and specifications for the two simulation suites used in this work.}
\label{tab:cosmo_sims}
\centering
\renewcommand{\arraystretch}{1.1}
\begin{tabular}{l c c}
\hline\hline
 & Abacus & Flagship 2 \\
\hline
   $\Omega_\mathrm{m}$   & 0.31519 & 0.319  \\
   $\Omega_\mathrm{b}$   & 0.0493  & 0.049  \\
   $h$                   & 0.6736  & 0.67   \\
   $10^9 \ A_\mathrm{s}$ & 2.083   & 2.1    \\
   $n_\mathrm{s}$        & 0.9649  & 0.96   \\
   $\sigma_8$            & 0.808   & 0.813  \\
   $m_\nu$ [eV]          & 0.06    & 0.06   \\
\hline
   $N_\mathrm{real}$ & 25 & 1 \\
   $L_\mathrm{box} \ [h^{-1} \ \mathrm{Mpc}]$ & 2~000 & 3~600 \\
   $N_\mathrm{part}$ & 6~192$^3$ & 16~000$^3$ \\
   $M_\mathrm{h}^\mathrm{min} \ [M_\odot/h]$ & $7.4 \times 10^{10}$ & $10^{10}$ \\
\hline
\end{tabular}
\end{table}

The Flagship 2 halo catalogue is populated with galaxies with the \textsc{SciPIC} code \citep{Scipic-Carretero+17,Scipic-Gonzalez+26}.
We refer the reader to \cite{Flagship-Castander+24} for all the details, while here we only summarise the main features.
Each halo is assigned a central galaxy, positioned at the halo centre and at rest with it.
The number of satellite galaxies per halo is assigned according to a power law with mass-scale $M_1=1.5\times 10^{11} \ M_\odot/h$ and slope $\alpha=1$.
Galaxy $r$-band luminosities are assigned via abundance matching (AM). The target luminosity function is a modified version of that in \cite{Dahlen+05}, based on multi-band GOODS data \citep{Giavalisco+04}, with a scatter of $\sigma_{\log L}=0.12$.
Galaxy colours and labels (red, green, blue) are assigned to each galaxy by fitting magnitude-dependent Gaussians to the low-redshift colour-magnitude diagram from the NYU-VAGC SDSS galaxy catalogue \citep{Blanton+05}.
These values are later used to assign each galaxy a spectral energy distribution from a series of templates and to compute emission line fluxes.
Satellite galaxies are assumed to be distributed according to a triaxial Navarro--Frenk--White (NFW) profile \citep{NFW} around halo centres.
To mimic colour segregation in high-mass haloes, the concentration parameter for green and blue galaxies is reduced by factors of 2 and 4, respectively.
Finally, satellite galaxy velocities are given by solving the Jeans equation in the dynamical equilibrium approximation.
Several additional galaxy properties, not used in the present analysis, are provided in \cite{Flagship-Castander+24}, including sizes, shapes, spectral information.
We accessed the galaxy sample through the CosmoHub platform \citep{CosmoHub}, a web-based service designed for efficient querying and distribution of large cosmological datasets.

\section{\scipiclite}
\label{sec:scipiclite}

In this section, we present a detailed description of the \scipiclite code, highlighting its overall structure and main capabilities.
This discussion is complemented by App.~\ref{app:scipiclite.features}, which illustrates some of the more advanced features of the code.

\scipiclite is designed to generate HOD mock galaxy catalogues from dark matter halo catalogues extracted from cosmological simulations, both in periodic boxes and light cones.
The library also provides pipelines to optimise the HOD parameters against target summary statistics.
\scipiclite builds upon the \textsc{SciPIC} code, retaining some of its core functionalities, but providing a more versatile tool with additional features for galaxy clustering studies.

\subsection{Central galaxies}
\label{subsec:scipiclite.centrals}

The HOD function for central galaxies represents the probability that a dark matter halo of a given mass hosts a galaxy in its centre.
This probability may vary in amplitude and shape for different galaxy populations.
In \scipiclite, central galaxies are probabilistically assigned via a Bernoulli trial with mean occupation $\mean{\Ncen}(\Mh)$, where $\Mh$ is the halo mass.
By default, \scipiclite places central galaxies in the centres of their host dark matter haloes and at rest with them (i.e. it assumes no velocity bias for central galaxies).
Among the occupation models natively implemented in \scipiclite, we describe below only those employed in this work.
The code, however, already includes a considerably larger set of parametrisations and supports user-defined HOD functional forms.

A typical choice for luminosity-limited samples is a central occupation function that rises from zero and reaches a plateau at high halo masses \citep[e.g.][]{Zheng2005_TheoreticalHOD}.
To describe this behaviour, we adopt the \textbf{Asymmetric Sigmoid} model: 
\begin{equation}
    \mean{\Ncen^\mathrm{AsymSig}} \ (\Mh) = \frac{\fmax}{ \left[1+\exp\left(-\frac{\deltalogMh}{\sigmalogM}+\ln(2^{1/\nu}-1)\right)\right]^\nu},
    \label{eq:scipiclite.centrals.asymmetricsigmoid}
\end{equation}
where, here and in what follows, $\deltalogMh=\logMh - \logMmin$, $M_{\rm min}$ is the characteristic mass of haloes hosting central galaxies, $\fmax$ denotes the incompleteness factor, $\sigmalogM$ sets the width of the transition, while $\nu$ controls its skewness, with larger values shifting the asymmetry towards high masses.
The normalisation of the exponential term ensures that the function reaches half its maximum value at $\deltalogMh = 0$.
For $\nu=1$, Eq.~\ref{eq:scipiclite.centrals.asymmetricsigmoid} reduces to the standard \textbf{Sigmoid} function.

Not all galaxy populations are expected to show a plateau in the central occupation at high halo masses.
For this reason, \scipiclite also implements several HOD models in which the central occupation declines in this regime.
Gaussian models have been used, for example, by \cite{DESI_ELG_HOD-Rocher+23}, and are implemented in \scipiclite.

However, following previous studies of ELGs in SAM \citep{SF_HOD-GonzalezPerez+18} and hydrodynamical simulations \citep{DESI-galaxy-halo_connection-Yuan+22}, which find an asymmetric central occupation with an extended tail towards high halo masses, we adopt a \textbf{LogNormal} HOD:

\begin{equation}
    \mean{\Ncen^\mathrm{LogNorm}} \ (\Mh) = \fmax \ \frac{e^{-\sigma_{\log M}^2/2}}{1+\deltalogMh} \ \exp\left[-\frac{\ln^2 (1+\deltalogMh)}{2\sigma_{\log M}^2}\right].
    \label{eq:scipiclite.centrals.lognormal}
\end{equation}

\scipiclite also implements a more complex \textbf{StarForming} HOD model, a slightly modified version from the one originally proposed by \cite{SF_HOD-Avila+20} for eBOSS ELGs:

\begin{equation}
    \mean{\Ncen^\mathrm{SF}} \ (\Mh) = 
        \begin{cases}
            \fmax \exp\left[-\frac{\left(\deltalogMh -\mathcal X\right)^2}{2\sigma^2_{\log M}}\right] & \text{if } \deltalogMh < 0, \\
            \fmax \exp\left[-\frac{\mathcal X^2}{2 \sigma^2_{\log M}}\right] \ 10^{\gamma \ \deltalogMh}  & \text{if } \deltalogMh \geq 0
        \end{cases},
    \label{eq:scipiclite.centrals.starforming}
\end{equation}

where $\mathcal X=\gamma\ln10 \ \sigma^2_{\log M}$.
With this expression, the distribution is continuous and differentiable at $\logMh=\logMmin$ and falls with slope $\gamma<0$ at high masses.

\begin{figure}
\centering
\includegraphics[width=\hsize]{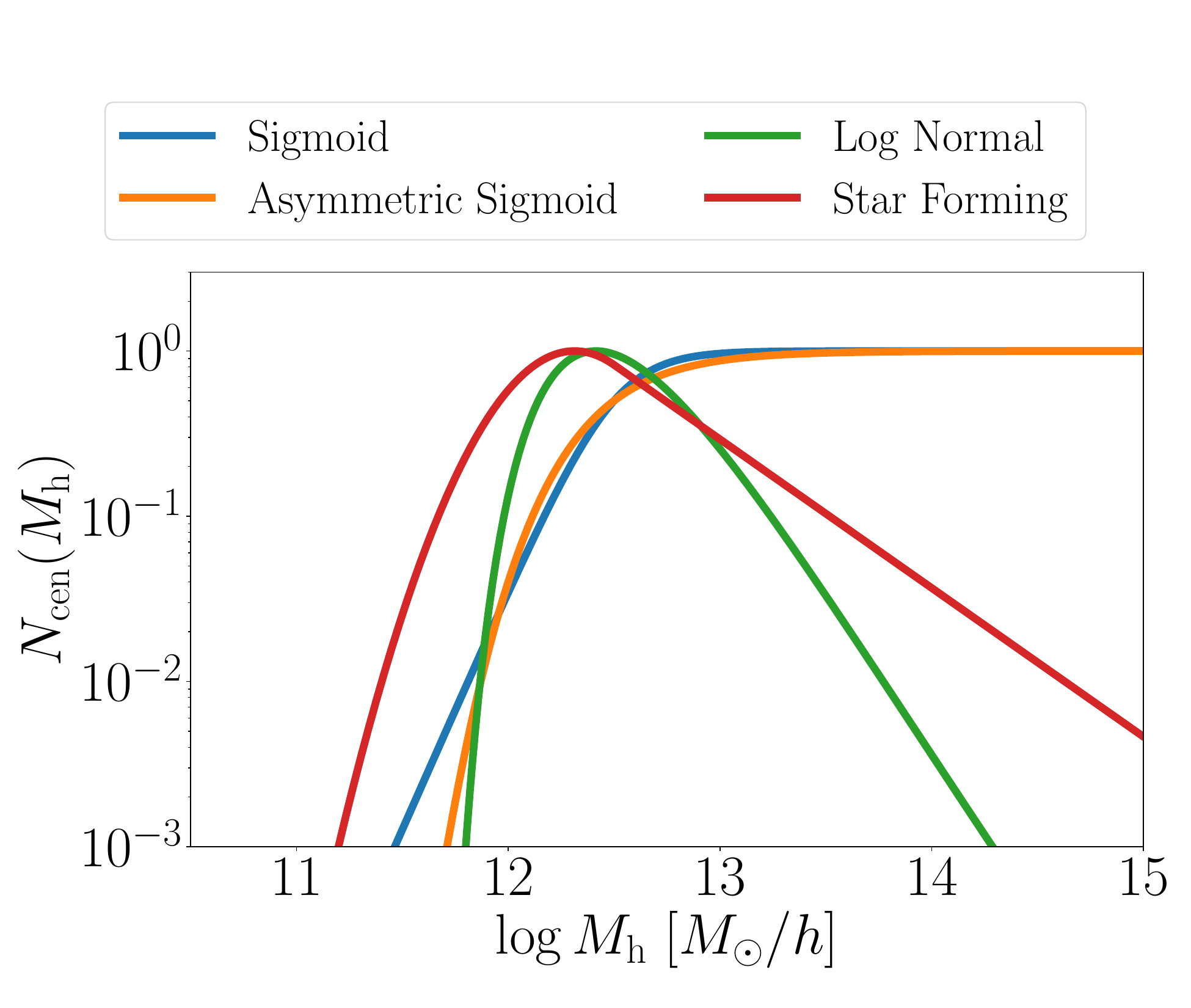}
     \caption{A sub-sample of HOD models for central galaxies currently implemented in \scipiclite, described in Sect. \ref{subsec:scipiclite.centrals} and used in this work. In orange and blue, an Asymmetric Sigmoid HOD (Eq.~\ref{eq:scipiclite.centrals.asymmetricsigmoid}) with its symmetric counterpart ($\nu=1$); in green a LogNormal HOD (Eq.~\ref{eq:scipiclite.centrals.lognormal}); in red a StarForming HOD (Eq.~\ref{eq:scipiclite.centrals.starforming}). We fix the values of the various parameters to $\fmax=1$, $\logMmin=12.5$, $\sigmalogM=0.15$ for Sigmoid and $\sigmalogM=0.3$ for the other models, $\gamma=-0.9$ for StarForming and $\nu=10$ for Asymmetric Sigmoid.}
     \label{fig:scipiclite.centrals.models}
\end{figure}

The HODs for central galaxies mentioned above can be visualised in Fig.~\ref{fig:scipiclite.centrals.models}.

\subsection{Satellite galaxies}
\label{subsec:hod.satellites}

In general, there is consensus on the expected mean number of satellite galaxies per halo, regardless of the galaxy population.
In this work, we always use the \textbf{Power Law} model:
\begin{equation}
    \mean{\Nsat^\mathrm{PowerLaw}} (\Mh) = \left(\frac{\Mh-\Mcut}{\Mone}\right)^\alpha,
    \label{eq:scipiclite.satellites.powerlaw}
\end{equation}

where $\Mcut$ is the cutoff mass, $\Mone$ is the halo mass scale for satellite galaxies and $\alpha$ is the slope of this relation.
Nevertheless, \scipiclite also implements other models for satellite occupation, e.g. a double power law.

\vspace{0.3cm}
\scipiclite places satellite galaxies following a given density profile around the halo centre.
By default, satellites are assumed to trace the dark matter distribution, which is commonly described by an NFW profile.
In \scipiclite, this profile is implemented in a generalised form with three free parameters:
\begin{equation}
    \rho_\ind{NFW}(r) \propto \frac{1}{\left(\frac{r}{R_\mathrm{s}}\right)^\alpha \ \left[1+\left(\frac{r}{R_\mathrm{s}}\right)^\gamma\right]^\beta},
    \label{eq:scipiclite.satellites.nfw}
\end{equation}
where $R_\mathrm{s}$ is the scale radius, while $\alpha$, $\gamma$, and $\beta$ control the inner slope, the sharpness of the transition, and the outer slope, respectively.
The standard NFW profile is recovered for $(\alpha, \beta, \gamma) = (1,2,1)$, which is the default configuration.
A generalised NFW profile was also employed in \cite{SAM_vs_HOD-Reyes-Peraza+23} to optimise the fit of the 2PCF based on SAM.

\scipiclite also implements the Einasto profile \citep{Einasto} and accepts any user-defined function of $r/R_\mathrm{s}$, allowing for arbitrary custom density profiles.
This makes it possible to adopt profiles calibrated on hydrodynamical simulations, or to account for baryonic effects on the inner structure of haloes \citep[e.g.][]{Baryonification_2015,Baryonification_I}, which can alter the spatial distribution of satellite galaxies.

\scipiclite places the galaxies within their host haloes assuming a triaxial geometry.
Each halo is characterised by its virial radius $R_\ind{vir}$, axis ratios and orientation, the latter defined through Euler angles.
The model also requires the concentration parameter
\begin{equation}
    c = \frac{R_\ind{vir}}{R_\ind{s}}.
    \label{eq:scipiclite.satellites.concentration}
\end{equation}

By default, galaxies are distributed out to a maximum radius of $3\,R_\ind{vir}$, and halo concentrations are rescaled by a factor of $1/4$, following the choices adopted by \citet{Flagship-Castander+24} for blue galaxies.
Both values can be adjusted by the user.

\vspace{0.3cm}
For simplicity, satellite galaxy velocities in \scipiclite are assigned under the assumption that haloes are spherical and dynamically relaxed.
Velocities are obtained by solving the Jeans equation.
In this approximation the relevant quantity is the radial velocity dispersion ($\sigma_r$), which satisfies
\begin{equation}
    \der{\sigma_r^2}{r} + \left[2\beta+\der{\ln\rho}{\ln r}\right]\frac{\sigma_r^2}{r} = - \frac{G M(<r)}{r^2},
    \label{eq:scipiclite.satellites.velocities}
\end{equation}

where $M(<r)$ is the halo mass enclosed within a radius $r$, $\rho$ is the halo density profile and $\beta=1-\sigma^2_\theta/\sigma_r^2$ is the anisotropy profile.

\scipiclite implements the \cite{velocity_anisotropy-Tiret+07} profile,
\begin{equation}
    \beta(r) = \beta_\infty \ \frac{r}{r+r_\ind{off} \ R_\mathrm{vir}},
    \label{eq:scipiclite.satellites.tiret}
\end{equation}

where $\beta_\infty$ and $r_\ind{off}$ are free parameters that can be tuned by the user.
Their default values are $0.82$ and $1$, respectively, as reported for the ``Blue Cloud'' population in Table~1 of \cite{Flagship-Castander+24}.
Like for position assignment, the concentration of the haloes is also globally rescaled by a factor of 1/4 by default (although it can be changed by the user).

In addition, \scipiclite provides the option to rescale the halo velocity dispersion to model \textit{velocity bias}, following an approach analogous to that of \cite{DESI_ELG_HOD-Rocher+23}.
A velocity-dispersion rescaling parameter, $b_v$, allows galaxy velocities to deviate systematically from those of the underlying dark matter.
This provides greater flexibility in modelling redshift-space clustering.
Finally, the procedure used to convert the radial velocity dispersion of the halo into three-dimensional galaxy velocities follows a methodology analogous to that described in Sect. 5.3 of \cite{Flagship-Castander+24}.

\subsection{HOD extensions}
\label{subsec:scipiclite.extensions}

In addition to the standard formalism, \scipiclite supports several extensions, including models with conformity between central and satellite galaxies, as well as models in which galaxy occupation depends on halo properties beyond mass.
In this section, we describe the ones employed in this work.
The remaining extensions are presented in App.~\ref{app:scipiclite.features}.

\paragraph{Conformity}
The standard HOD framework treats satellite occupation as independent of the central galaxy, thereby neglecting 1-halo \textit{conformity}, namely the tendency of galaxies in the same halo to share properties beyond those set by halo mass.
Hydrodynamical simulations such as IllustrisTNG indicate that Luminous Red Galaxies (LRGs) exhibit little to no conformity, since they reside almost exclusively in massive haloes where satellites are common \citep{Conformity_Illustris-Yuan+25}.
In contrast, comparisons between DESI observations and HOD-based mocks suggest strong 1-halo conformity for ELGs \citep{DESI_ELG_HOD-Rocher+23,DESI_ELG_conformity-Gao+23}.
\citet{Assembly_bias_hydro-occupation-Hadzhiyska+23a} found that haloes hosting an ELG satellite are nearly twice as likely to host an ELG central, and accounted for this effect through an additional free parameter in the HOD.

\scipiclite adopts this latter approach, focusing on the impact of 1-halo conformity on the satellite occupation.
The mean number of satellites is rescaled according to whether the halo hosts a central galaxy:
\begin{eqnarray}
    \mean{N_\ind{sat} \ \left| \ N_\ind{cen} = 1 \right.} &=& C_+ \cdot \mean{N_\ind{sat}}
    \nonumber
    \\
    \mean{N_\ind{sat} \ \left| \ N_\ind{cen} = 0 \right.} &=& C_- \cdot  \mean{N_\ind{sat}},
    \label{eq:scipiclite.conformity}
\end{eqnarray}
where $C_+$ and $C_-$ can be provided either as scalars or as functions of halo mass.

\paragraph{Assembly bias}
The dependence of galaxy occupation on secondary halo properties beyond mass, known as galaxy \textit{assembly bias}, is supported by growing evidence from observations \citep{Assembly_bias_SDSS-Wang+22,Assembly_bias_CIC_DESI-Pearl+23,Assembly_bias_eBOSS-Paviot+24,DESI_ELG_HOD-Rocher+23} and hydrodynamical simulations \citep{Assembly_bias_hydro-Chaves-Montero+16,Assembly_bias_hydro-Artale+18,Assembly_bias_hydro-Garcia-Moreno+25}.
Halo formation history and environment, in particular, have been shown to affect galaxy clustering \citep{Assembly_bias_hydro-Bose+19,Assembly_bias_hydro-Hadzhiyska+20,Assembly_bias_hydro-Zehavi+18,Assembly_bias_hydro-occupation-Hadzhiyska+23a,Assembly_bias_hydro-occupation-Hadzhiyska+23b}, although no single property appears sufficient to account for the full signal \citep{Assembly_bias_hydro-Garcia-Moreno+25}.
The assembly bias amplitude can vary depending on the galaxy population.
\cite{Assembly_bias_eBOSS-Paviot+24} found strong evidence that LRG centrals in eBOSS mostly occupy denser environments.
By contrast, ELGs are less affected by secondary biases, a result that was also found by \cite{DESI_ELG_HOD-Rocher+23}.

In \scipiclite, assembly bias is incorporated by introducing shifts in key HOD parameters for individual haloes - typically $\log M_\mathrm{min}$ for centrals and $\log M_1$ for satellites, since these parameters predominantly determine the clustering amplitude.
By adjusting these parameters according to secondary halo properties, the model can capture how halo assembly history influences galaxy occupation and large-scale clustering.
Specifically, haloes are first grouped into narrow mass bins and then ranked according to a secondary property $\Theta$, with values between $-1$ and $1$.
Additional free parameters (one for each property) are introduced to control the amplitude of the assembly bias effect\footnote{Note that the same procedure can be exploited to introduce e.g. redshift-dependent HOD parameters, see App.~\ref{app:scipiclite.features}.}.
In formulae, given a halo $h$ and a series of secondary properties $\Theta$, the HOD parameters transform on a halo-by-halo basis according to:
\begin{eqnarray}
    \logMmin &\rightarrow& \logMmin + \sum_\Theta A_\mathrm{cen,\Theta} \ \mathrm{Rank}\left[\Theta_{h}\right]
    \nonumber
    \\
    \logMone &\rightarrow& \logMone + \sum_\Theta  A_\mathrm{sat,\Theta} \  \mathrm{Rank}\left[\Theta_{h}\right] \ .
    \label{eq:scipiclite.assembly_bias}
\end{eqnarray}

\subsection{Clustering tools}
\label{subsec:scipiclite.clustering}

\scipiclite includes a wrapper around the \texttt{Corrfunc} software \citep{Corrfunc}, which offers greater flexibility in the measurement of 2PCFs with different estimators.
In this work, we employ the Landy--Szalay estimator \citep{LandySzalay},
\begin{equation}
    \xi = \frac{DD-2DR+RR}{RR},
    \label{eq:scipiclite.clustering.LS_estimator}
\end{equation}
and the natural estimator \citep{Peebles1974},
\begin{equation}
    \xi = \frac{DD}{RR}-1,
    \label{eq:scipiclite.clustering.natural_estimator}
\end{equation}
where $DD$, $DR$, and $RR$ are the normalised data--data, data--random, and random--random pair counts.
Anisotropic 2PCFs can be measured either in bins of separation perpendicular and parallel to the line of sight, $(r_\mathrm{p}, r_\pi)$, or in bins of total separation and orientation, $(s, \mu)$.
From these measurements, the code derives the projected 2PCF and the multipoles of the 2PCF,
\begin{eqnarray}
w_\mathrm{p}(r_\mathrm{p}) &=& 2 \int_0^{r_{\pi,\mathrm{max}}} \de r_\pi \ \xi(r_\mathrm{p},r_\pi),
\label{eq:scipiclite.clustering.projected_2PCF}
\\
\xi_\ell(s) &=& \frac{2\ell+1}{2} \int_{-1}^{1} \de\mu \ \xi(s,\mu) \ \mathcal{L}_\ell(\mu),
\label{eq:scipiclite.clustering.2PCF_multipoles}
\end{eqnarray}
where $\mathcal{L}_\ell$ is the Legendre polynomial of order $\ell$.

In addition to the 2PCF, \scipiclite natively implements routines to measure angular 2PCFs, density fields, power spectra, and excess surface density profiles.

\subsection{Minimisation pipeline}
\label{subsec:scipiclite.pipeline}

\scipiclite provides a complete framework for parameter inference on HOD models, with ready-to-use Python notebooks combining halo population, clustering measurements, and likelihood sampling within a single workflow.

Parameter inference generally follows one of two approaches.
The first is the direct minimisation of a $\chi^2$ function, which identifies the best-fit set of parameters.
The second is the full exploration of the posterior distribution, typically through Markov Chain Monte Carlo (MCMC) methods.
Both approaches rely on a likelihood or a $\chi^2$ function,
\begin{equation}
    \chi^2(\boldsymbol{\Theta}) = \sum_{i,j} \left[\boldsymbol{d}-\boldsymbol{m}(\boldsymbol{\Theta})\right]_i \ \left[\mathrm{Cov}^{-1}\right]_{ij} \
    \left[\boldsymbol{d}-\boldsymbol{m}(\boldsymbol{\Theta})\right]_j,
    \label{eq:scipiclite.pipeline.chi2}
\end{equation}
where $\boldsymbol{d}$ is the data vector, $\boldsymbol{m}(\boldsymbol{\Theta})$ is the model evaluated at the parameters $\boldsymbol{\Theta}$, and Cov is the covariance matrix.

In the case of HOD fitting, each approach faces a specific difficulty.
Direct minimisation is hindered by the stochastic nature of the HOD model, which arises from the random assignment of galaxies to haloes.
Classical gradient-based minimisers are therefore unreliable.
Posterior sampling, on the other hand, is limited by the computational cost of each model evaluation, which requires populating the haloes and measuring the 2PCF.
Standard MCMC methods, which typically demand a very large number of evaluations, thus become impractical.
\scipiclite implements a dedicated solution for each of these limitations.

For direct minimisation, \scipiclite adopts \textit{differential evolution} \citep{Differential_evolution-Storn-Price+97}, a stochastic minimiser well suited to noisy problems.
This method has already been applied in similar contexts \citep{DES_yr3-Ferrero+21,Stochastic_minimizer-Tutusaus+25}.
Differential evolution iteratively evolves a population of candidate solutions, generating new trial points by combining the differences between randomly selected members.
A trial point replaces its parent only if it yields a smaller $\chi^2$, so that the population progressively converges towards the global minimum.

For posterior sampling, \scipiclite implements an iterative emulator-based approach, first introduced in a cosmological context by \citet{Iterative_emulator-Pellejero+19} and later adopted in several HOD studies \citep[e.g.][]{Rocher_GP,DESI_ELG_HOD-Rocher+23}.
In this approach, the $\chi^2$ function is emulated from a sparse sampling of the parameter space and then explored with MCMC methods.
The emulator is refined iteratively, with new evaluations concentrated in the regions of highest posterior probability.
This reduces the number of explicit likelihood evaluations by up to a factor of $\sim100$, without compromising the accuracy of the recovered posterior.

Python notebooks demonstrating both procedures on a custom 2PCF are provided in the \scipiclite repository.
In this work, we adopt differential evolution for the main analysis, while an illustrative example of the iterative MCMC procedure is presented in App.~\ref{app:scipiclite.iterative_fit}.

\subsection{Performance}
\label{subsec:scipiclite.performance}

We assess the computational performance of \scipiclite by running a single evaluation of the full fitting pipeline, from mock generation to the computation of the likelihood.
We use our fiducial halo catalogue mentioned in Sect. \ref{subsec:simulations.abacus}, focusing on the snapshot at $z=0.95$ that contains approximately 400 million objects.

Galaxies are placed into haloes using an Asymmetric Sigmoid occupation function for central galaxies (Eq.~\ref{eq:scipiclite.centrals.asymmetricsigmoid}) and a Power Law for satellites.
The HOD parameters are fixed by requiring a galaxy number density of $n_\mathrm{gal} = 1.43 \times 10^{-3}\,(\hMpc)^{-3}$ (see Table \ref{tab:FS2_stats}) and yield a mock catalogue of about 11.4 million galaxies.
From this catalogue, we measure the projected 2PCF and the 2PCF multipoles, using the natural estimator with analytical random pair counts for the periodic box.

Mock generation requires on average $19.1\,\mathrm{s}$ of wall-clock time on a single CPU core, as estimated over 50 realisations.
The construction of the galaxy catalogue takes $6.2\,\mathrm{s}$; the assignment of satellite positions takes a comparable $8.8\,\mathrm{s}$; the assignment of satellite velocities requires $3.3\,\mathrm{s}$, while the mapping from real to redshift space requires only $0.8\,\mathrm{s}$.
Pair counting dominates the computational cost of the pipeline, requiring $47.1\,\mathrm{s}$ to measure the projected 2PCF and the multipoles up to a maximum separation of $30\,h^{-1}\mathrm{Mpc}$.
The time required to compute the $\chi^2$ is negligible.
Both the population of haloes and the pair counting are suitable for parallelisation, which would substantially reduce the cost of each likelihood evaluation.

\section{Calibration dataset}
\label{sec:dataset}

\begin{figure}
    \centering
    \includegraphics[width=\hsize]{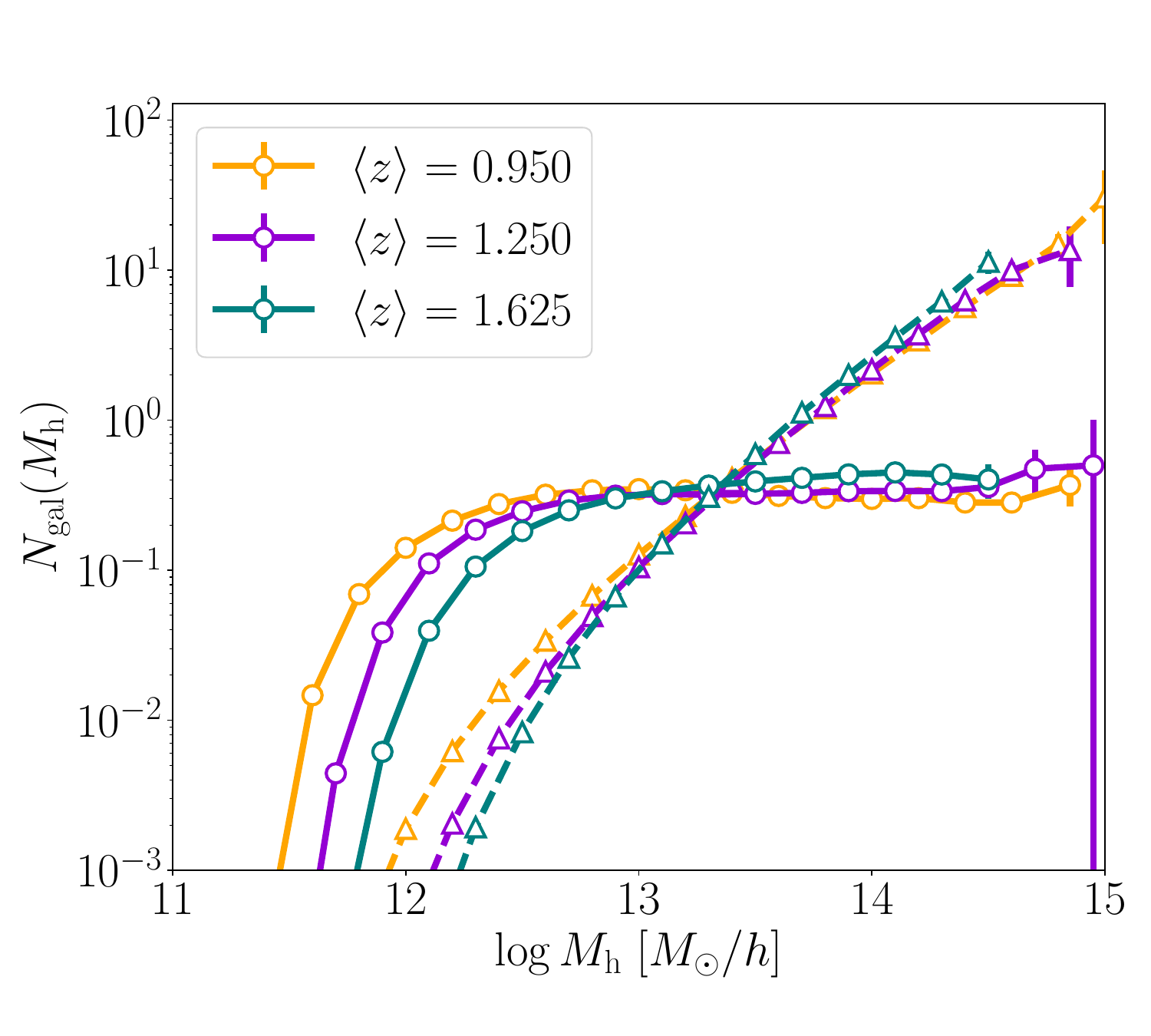}
    \caption{HOD of the Flagship~2 H$\alpha$ galaxy sample, as measured by directly counting the number of galaxies per halo in the fiducial footprint, for the three redshift bins used in this work. Solid lines and circles represent the occupation function for centrals, while dashed lines and triangles correspond to satellites. Note that symbols include Poisson error bars.}
    \label{fig:FS2_HOD}
\end{figure}

In this section, we describe the data used to validate the \scipiclite HOD fitting pipeline.
Our fiducial dataset consists of two-point statistics measured from a sub-sample of the Flagship~2 galaxy catalogue (Sect.~\ref{subsec:simulations.flagship2}), which we fit to constrain the parameters of the HOD models under consideration.
We start by cutting from the Flagship 2 octant a circular footprint of area 19.3 deg$^2$.
We will refer to this as the \textit{fiducial footprint}.
We then select, in that pencil-like cone, Flagship 2 galaxies with an H$\alpha$ flux larger than $f_{{\rm H}\alpha, \mathrm{min}} = 2 \times 10^{-16} \ \mathrm{erg \ s^{-1} \ cm^{-2}}$ in 3 redshift bins with edges $z=[(0.85,1.04),(1.10,1.39),(1.45,1.80)]$.
This choice is driven by the need to match the redshift range of the expected \textit{Euclid} H$\alpha$ galaxies, while also ensuring consistency in the mean halo redshift between the Flagship 2 light cone and Abacus snapshots.
The fluxes that we consider include internal attenuation and have been previously renormalised in \cite{Flagship-Castander+24} in order to match the number density predictions for H$\alpha$ galaxies in \cite{Pozzetti+16}.
In Table \ref{tab:FS2_stats}, we present a set of useful summary statistics for our galaxy sample.
We show in Fig.~\ref{fig:FS2_HOD} the resulting number of central and satellite galaxies per halo as directly measured from the Flagship 2 simulation.
Given its resemblance to an asymmetric sigmoid function, we adopt Eq.~\ref{eq:scipiclite.centrals.asymmetricsigmoid} as our fiducial HOD model throughout this work.

\begin{table}
\caption{Properties of the Flagship~2 (FS2 here) H$\alpha$ galaxy sample used in this work. The minimum flux has been set to $f_{{\rm H}\alpha, \mathrm{min}} = 2 \times 10^{-16} \ \mathrm{erg \ s^{-1} \ cm^{-2}}$.}
\label{tab:FS2_stats}
\centering
\renewcommand{\arraystretch}{1.1}
\begin{tabular}{c c c c}
\hline\hline
$\mean{z_\mathrm{Abacus}}$ & $z_\mathrm{FS2}$ range & 
\begin{tabular}{c}
$\bar n_\mathrm g $ \\
$[h^{3} \ \rm{Mpc}^{-3}]$
\end{tabular}  &
\begin{tabular}{c}
Volume \\
{$[h^{-3} \ \rm{Mpc}^3]$}
\end{tabular} \\
\hline
   0.950 & 0.85-1.04 & $1.43 \times 10^{-3}$ & $9.5 \times 10^6$ \\
   1.250 & 1.10-1.39 & $7.63 \times 10^{-4}$ & $1.7 \times 10^7$ \\
   1.625 & 1.45-1.80 & $3.72 \times 10^{-4}$ & $2.4 \times 10^7$ \\
\hline
\end{tabular}
\end{table}

For each redshift bin, we first generate a random catalogue containing 30 times as many objects, matched to the same angular footprint and redshift distribution.
We then compute the projected 2PCF and the 2PCF multipoles using the Landy–Szalay estimator (Eq.~\ref{eq:scipiclite.clustering.LS_estimator}).
For the projected 2PCF, we adopt 30 logarithmic bins in $r_\mathrm{p}$ from 0.1 to 30 $\hMpc$; the integration along the line of sight is performed up to $r_{\pi,\mathrm{max}}=40 \ \hMpc$.
For the redshift-space multipoles, we use 30 logarithmic bins in $s$ from 0.3 to 30 $\hMpc$.
In this work, we use the monopole ($\ell=0$) and the quadrupole ($\ell=2$).

\begin{figure}[t!]
    \centering
    \includegraphics[width=\columnwidth]{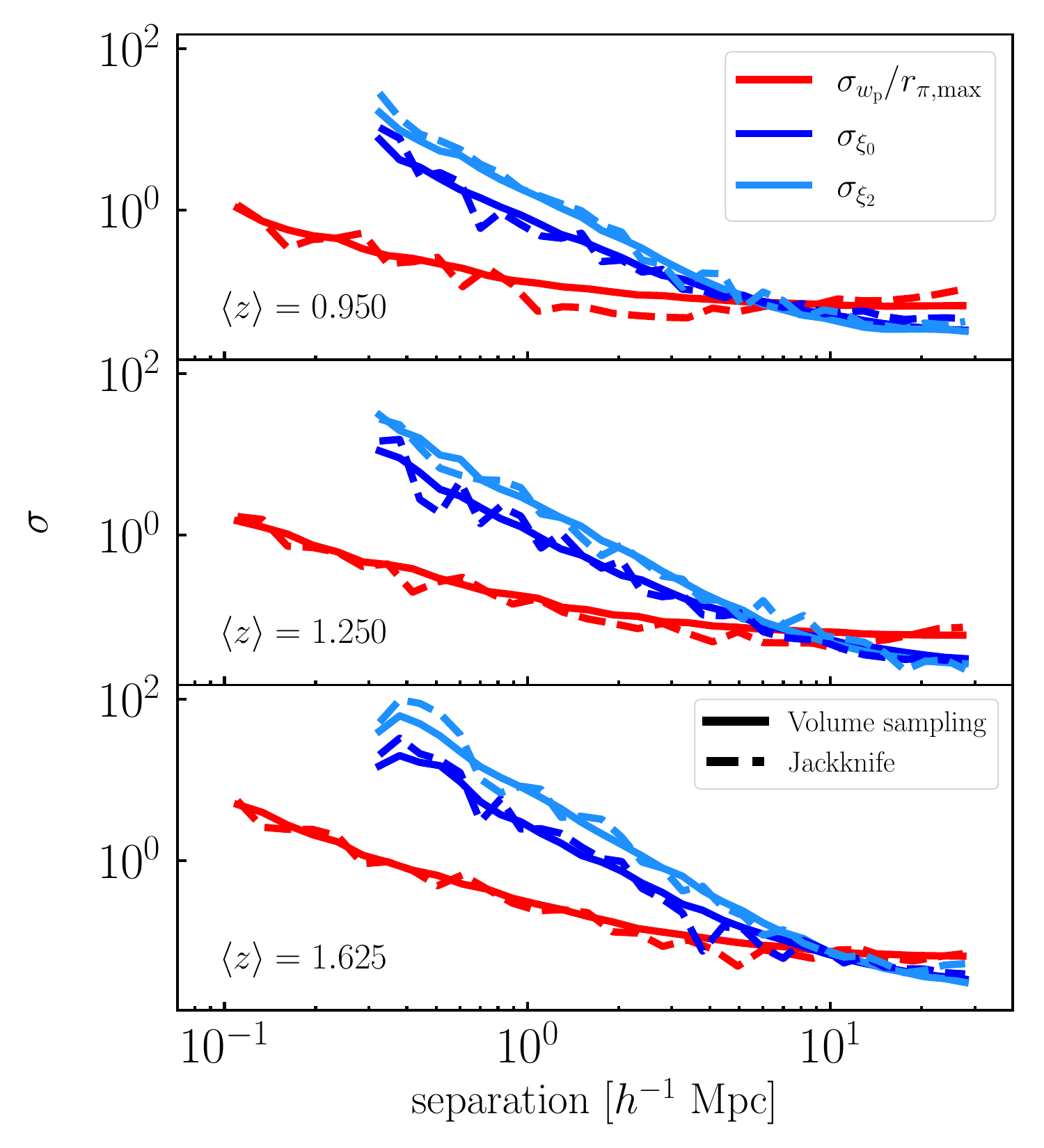}
    \caption{Comparison between the statistical uncertainties of the measured 2PCFs, estimated from the diagonal of the covariance matrix based on resampling regions of the Flagship 2 octant (solid lines), and those obtained via jackknife resampling of the fiducial footprint (dashed lines). Each panel shows a different redshift slice used in this work. The $x$-axis represents $r_\mathrm{p}$ or $s$, depending on the observable shown. In red we show the uncertainties on the projected 2PCF (rescaled by $r_{\pi,\mathrm{max}}$ to improve the readability of the figure), in blue and light blue those on the monopole and quadrupole of the 2PCF, respectively.}
    \label{fig:FS2_variance}
\end{figure}

We exploit the large volume of the Flagship 2 simulation to estimate the covariance matrix of our clustering observables.
To this end, we apply random rotations to the Flagship 2 octant and overlay our fiducial footprint onto the rotated volume.
We repeat this procedure 2000 times, obtaining an ensemble of measurements large enough to build a smooth and well-converged covariance matrix.
To speed up the computation, we use the natural estimator (Eq.~\ref{eq:scipiclite.clustering.natural_estimator}).
Since the sky area of Flagship 2 is only $\sim 250$ times larger than our footprint, some regions are inevitably reused across realisations.
To test whether this repetition affects our covariance estimate, we also compute a jackknife estimate. Given the angular extent of our footprint and the maximum separation considered, at most 10 independent regions can be defined. This number is far too small to estimate the off-diagonal elements of the covariance matrix reliably, so we restrict the comparison to the variance.
The results are shown in Fig.~\ref{fig:FS2_variance}.
For both the projected 2PCF and the 2PCF multipoles, the two methods agree remarkably well in all redshift bins.
We therefore adopt this covariance matrix estimate throughout this work.

\section{Results}
\label{sec:results}

\begin{table*}[h!]
\caption{HOD parameters used in this work, their meaning and their priors. The parameters have been split according to their role in the HODs and in their extensions. The label $\mathcal U$ indicates a uniform prior between the two extrema in square brackets.}
\label{tab:priors}
\centering
\renewcommand{\arraystretch}{1.1}
\begin{tabular}{l l c c }
\hline\hline
\textbf{Parameter} & \textbf{Meaning} & \textbf{Prior} & \textbf{Default value} \\
\hline
\multicolumn{4}{c}{\rule{0pt}{12pt}\rule[-6pt]{0pt}{0pt}\textit{Baseline HOD}} \\
\hline
   $\logMmin$   & Halo mass scale for hosting a central galaxy     & $\mathcal U[11.6,13.5]$ & $-$ \\
   $\sigmalogM$ & Transition width for central galaxies            & $\mathcal U[0.01,0.7]$ & $-$ \\
   $\log\nu$    & Asymmetry of the sigmoid function                    & $\mathcal U[0,2]$ & $-$ \\
   $\gamma$     & High-mass star formation quenching parameter     & $\mathcal{U}[-2,0]$ & $-$ \\
   $\logMcut$   & Halo cutoff mass for hosting a satellite galaxy & $\mathcal U[11.6,13.0]$ & $-$ \\
   $\logMone$   & Halo mass scale for hosting a satellite galaxy   & $\mathcal U[13,15]$ & $-$ \\
   $\alpha$     & Exponent of satellite power law                  & $\mathcal U[0.3,2.5]$ & $-$ \\
\hline
\multicolumn{4}{c}{\rule{0pt}{12pt}\rule[-6pt]{0pt}{0pt}\textit{Profile rescaling}} \\
\hline
   $R_\mathrm{max}$  & Maximum radius to place satellites in $R_\mathrm{vir}$ units & $\mathcal U[1,4]$ & 3 \\
   $c_\mathrm{norm}$ & Concentration rescaling parameter  & $\mathcal U[1,5]$ & 4 \\
\hline
\multicolumn{4}{c}{\rule{0pt}{12pt}\rule[-6pt]{0pt}{0pt}\textit{Assembly bias}} \\
\hline
   $A_\mathrm{cen}$ & Assembly bias parameter for central galaxies   & $\mathcal U[-1,1]$ & 0 \\
   $A_\mathrm{sat}$ & Assembly bias parameter for satellite galaxies & $\mathcal U[-1,1]$ & 0 \\
\hline
\multicolumn{4}{c}{\rule{0pt}{12pt}\rule[-6pt]{0pt}{0pt}\textit{Conformity}} \\
\hline
   $C_+$ & Rescaling of the number of satellites with a central galaxy    & $\mathcal U[0.5,2]$ & 1 \\
   $C_-$ & Rescaling of the number of satellites without a central galaxy & $\mathcal U[0.5,2]$ & 1 \\
\hline
\multicolumn{4}{c}{\rule{0pt}{12pt}\rule[-6pt]{0pt}{0pt}\textit{Velocity bias}} \\
\hline
   $b_v$ & Rescaling of satellite velocity dispersion    & $\mathcal U[0.5,2]$ & 1 \\
\hline
\end{tabular}
\end{table*}

In this section, we present the results of fitting the 2PCFs measured in Sect.~\ref{sec:dataset}, following the procedure described in Sect.~\ref{subsec:scipiclite.pipeline}.
Our aim is to recover the clustering of the Flagship~2 H$\alpha$ sample and to assess the constraining power of different clustering statistics, namely the projected 2PCF, the redshift-space multipoles, and the joint analysis.
We compare the measurements with the predictions of the best-fit models, considering both standard HOD functional forms and several physically motivated extensions, aimed at improving the modelling of the clustering signal at small separations.

\subsection{Modelling choices and halo catalogue pre-processing}
\label{subsec:results.models}

We test different HOD models: the Asymmetric Sigmoid model (our fiducial model, Eq.~\ref{eq:scipiclite.centrals.asymmetricsigmoid}), the Sigmoid HOD, the LogNormal HOD (Eq.~\ref{eq:scipiclite.centrals.lognormal}) and the StarForming HOD (Eq.~\ref{eq:scipiclite.centrals.starforming}).
For satellite galaxies, we always assume the Power Law function (Eq.~\ref{eq:scipiclite.satellites.powerlaw}). 
We always assume that satellite occupation follows Poisson statistics and that satellites are distributed around halo centres according to an NFW profile.
We also consider the following extensions:
\begin{itemize}
    \item \textit{Profile rescaling}: in our baseline model, satellites populate haloes out to $R_\mathrm{max} = 3 \ R_\mathrm{vir}$, and halo concentrations are rescaled by a factor $c_\mathrm{norm} = 4$.
    In this extension, both are treated as free parameters.

    \item \textit{Assembly bias}: two parameters ($A_\mathrm{cen}$, $A_\mathrm{sat}$) modulate the $\logMmin$ and $\logMone$ parameters, respectively, on a halo-by-halo basis following Eq.~\ref{eq:scipiclite.assembly_bias} to model central and satellite assembly bias.
    As the secondary halo property beyond mass, we adopt the local halo overdensity, interpolated on a grid of size $512^3$ and then evaluated at halo positions.

    \item \textit{Conformity}: two parameters ($C_+$, $C_-$) modulate the satellite occupation function depending on the presence or absence of a central galaxy in the same halo (Eq.~\ref{eq:scipiclite.conformity}). The default values in the standard HOD picture are $C_+=C_-=1$.

    \item \textit{Velocity bias}: satellite velocity dispersions (Eq.~\ref{eq:scipiclite.satellites.velocities}) are rescaled by a factor $b_v$ to model discrepancies in the galaxy vs. dark matter velocity field. The default value for the velocity bias parameter is 1.
\end{itemize}
The parameters and priors adopted in all analyses are summarised in Table~\ref{tab:priors}.

As mentioned in Sect. \ref{subsec:simulations.abacus}, our ``modelling set'' consists of the halo catalogue of the first realisation of the Abacus simulation with Planck cosmology.
We refrain from populating the Flagship~2 haloes directly, in order to avoid the additional complications introduced by their light-cone geometry.
In a light-cone configuration, each set of HOD parameters yields a different redshift distribution, $n(z)$, of the mock galaxies.
A consistent clustering analysis would then require generating, at every step of the sampling, a random catalogue that follows the corresponding $n(z)$, and re-measuring the random--random pair counts accordingly, thereby increasing the computational cost of each likelihood evaluation by a large factor.
Although \scipiclite\ is already equipped to perform this procedure, its implementation lies beyond the scope of the present work.
We discuss some considerations that need to be taken into account and outline the adjustments required to optimise the methodology.

\begin{itemize}
    \item The cosmologies of Abacus and Flagship~2 differ slightly (see Table \ref{tab:cosmo_sims}).
    However, the resulting differences in the linear growth factors remain below 0.1\% in all redshift bins, and we therefore neglect them.

    \item The modelling volume must be considerably larger than the measurement volume, in order to minimise the impact of cosmic variance and model stochasticity.
    A single Abacus snapshot covers a volume between 330 and 840 times larger than our fiducial footprint, depending on redshift.

    \item The different halo finders adopted in Abacus and Flagship~2 lead to discrepancies of up to $\sim 20\%$ in the halo mass function at the high-mass end, with a significant impact on halo bias.
    Although these differences would be absorbed during the fit, we apply abundance matching to the Abacus halo masses so that their mass function matches that of Flagship~2, to enable a direct comparison between their halo populations.

    \item The CompaSO halo finder does not provide scale or virial radii (Eq.~\ref{eq:scipiclite.satellites.concentration}), but only radii enclosing fixed fractions of the halo mass.
    These quantities are needed to estimate the halo concentration parameter, which in turn sets how satellites are distributed within haloes, and hence the clustering on small scales.
    Previous works \citep[e.g.][]{AbacusHOD-Yuan+21,DESI_ELG_HOD-Rocher+23} estimated halo concentrations from the ratio between either $r_{90}$ or $r_{98}$ and $r_{25}$.
    However, we find that the resulting concentration--mass relation is almost flat and less scattered than in Flagship~2, which may affect the satellite distribution and the small-scale clustering.
    We therefore bin the Flagship~2 haloes in mass and measure the cumulative distribution function (CDF) of their concentrations in each bin.
    Concentrations are then reassigned to the Abacus haloes by sampling from this CDF, while preserving their original rank ordering and thus any pre-existing correlations.

    \item The number density constraint allows one parameter to be derived from the others.
    We choose $\fmax$ as the dependent parameter, obtained by imposing
    \begin{equation}
        \begin{split}
            \bar n^\mathrm{survey}_\mathrm g = {} & \int \de \Mh \ \frac{\de n_\mathrm h}{\de \Mh}
            \int_{-1}^{1} \de \Theta \ \mathcal P(\Theta|\Mh) \\
            & \times \Bigg\{\mean{\Ncen}+\mean{\Nsat} \ \Big[\mean{\Ncen} \ C_+ + \big(1-\mean{\Ncen}\big) \ C_-\Big]\Bigg\}.
        \end{split}
        \label{eq:results.number_density}
    \end{equation}

    In the equation above $\frac{\de n_\mathrm h}{\de \Mh}$ is the halo mass function, $\mathcal P(\Theta|\Mh)$ is the PDF of a secondary property $\Theta$ that encodes assembly bias, $C_\pm$ are the conformity parameters and $\mean{\Ncen}$ and $\mean{\Nsat}$ actually depend on both $\Mh$ and $\Theta$\footnote{The standard HOD prediction is recovered when $C_\pm=1$, $\mathcal P=1/2$ and $\mean{\Ncen}$ and $\mean{\Nsat}$ only depend on halo mass.}.

    \item At each evaluation of the model, the 2PCFs are measured using the natural estimator (Eq.~\ref{eq:scipiclite.clustering.natural_estimator}), with the $RR$ pairs computed analytically by exploiting the periodicity of the simulation box.
\end{itemize}

\subsection{Baseline HOD}
\label{subsec:results.baseline}

\begin{figure*}[h!]
    \centering
    \includegraphics[width=0.9\textwidth]{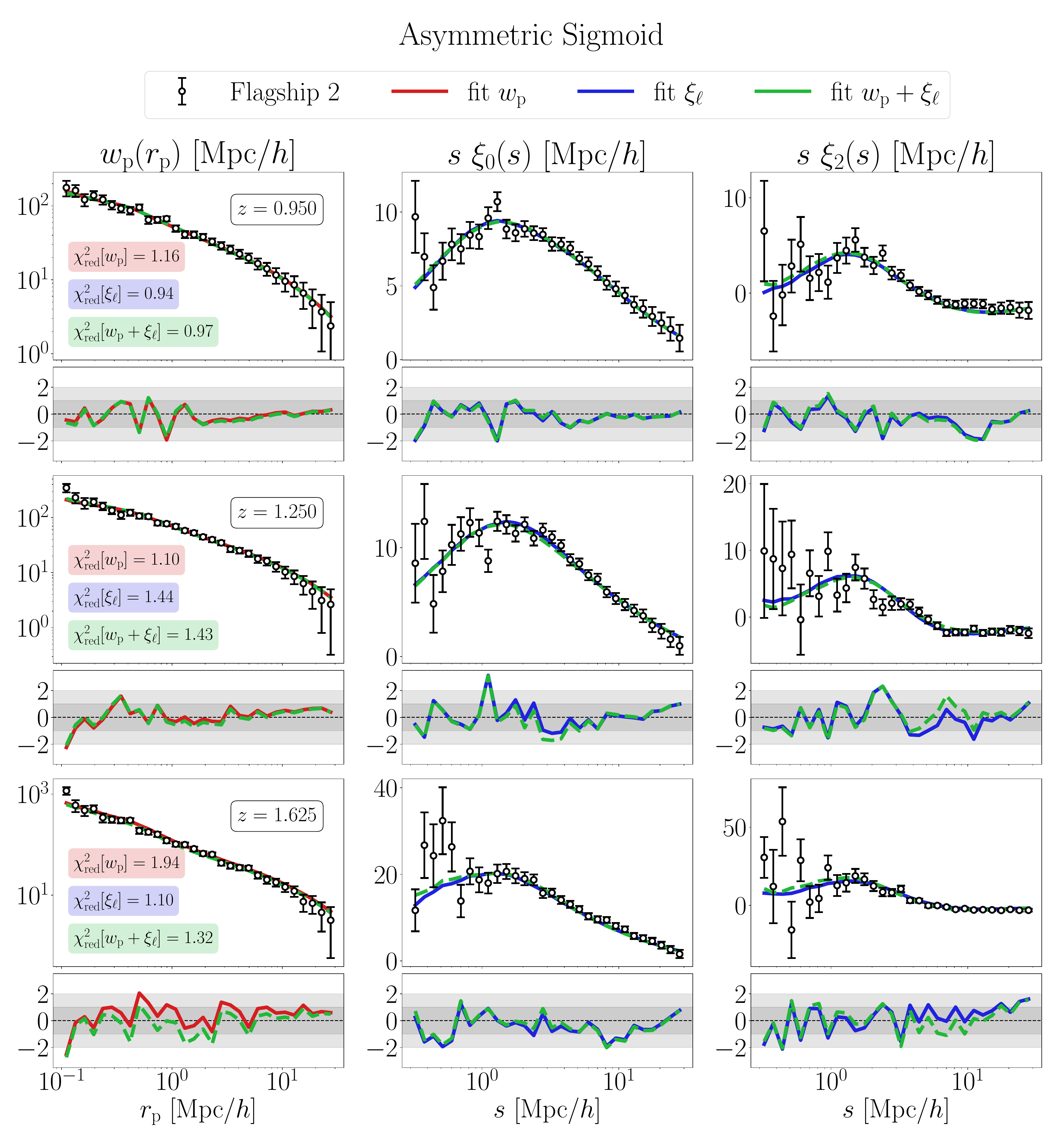}
    \caption{Best-fit 2PCF models for the Asymmetric Sigmoid parametrisation across the three redshift bins ($z = 0.950$, $1.250$, and $1.625$). Left, centre and right columns show the projected 2PCF, the 2PCF monopole and the 2PCF quadrupole, respectively. Data points from the Flagship~2 simulation are shown as black circles. Solid dark red, solid blue, and dashed green lines correspond to the best-fit models obtained by fitting the projected 2PCF only ($w_\mathrm{p}$), the 2PCF multipoles only ($\xi_\ell$), and both statistics jointly, respectively. The reduced chi-squared values $\redchisq$ for each fitting configuration are reported in each $w_\mathrm{p}$ panel. Lower sub-panels display the residuals between the data and the best-fit models, expressed in units of standard deviation.}

    \label{fig:FS2_fit_redshifts}
\end{figure*}

We begin by presenting the results of our minimisation process, keeping the HOD fixed to our fiducial model, the Asymmetric Sigmoid of Eq.~\ref{eq:scipiclite.centrals.asymmetricsigmoid}, and displaying the best fits in each redshift bin.
Fig.~\ref{fig:FS2_fit_redshifts} compares the best-fit models obtained from fitting $w_\mathrm{p}$ alone, the multipoles alone, and both statistics jointly, in the three redshift bins.
For all three configurations we report the values of the corresponding reduced $\chi^2$ (hereafter denoted $\redchisq$).
This is obtained by dividing Eq.~\ref{eq:scipiclite.pipeline.chi2} by the number of degrees of freedom, defined as the number of data points minus the number of free parameters.

A visual inspection of the figure reveals that the best-fit model is largely robust across the different observables considered.
In the joint fit, the $\redchisq$ values are driven primarily by the 2PCF multipoles $\xi_\ell$, which dominate the fit particularly in the modelling of the 1-halo term, where in-halo velocities play a more significant role than in the projected 2PCF.
As expected, the joint combination $\wprp+\xiell$ is the most robust probe, since it carries complementary information:
$\wprp$ integrates out redshift-space distortions and isolates the real-space clustering, while $\xiell$ is sensitive to both small-scale in-halo dynamics and large-scale halo bias.
Combining the two breaks degeneracies between parameters that remain poorly constrained when each probe is fitted on its own.
This holds both when minimising the $\chi^2$ and when sampling the full posterior with MCMC, where the joint fit gives noticeably tighter constraints on the model parameters (see App.~\ref{app:scipiclite.iterative_fit}).

The highest redshift bin proved to be the most challenging to model.
The $\redchisq$ obtained from fitting the projected 2PCF $w_\mathrm{p}$ alone is notably large ($\redchisq = 1.94$), reflecting a poor description of the data particularly at small scales, where the 1-halo term dominates.
We attribute this discrepancy to the treatment of satellite positions within haloes: in our pipeline, the concentration parameter is renormalised by a factor of 4 when placing satellites, an operation that was originally performed in Flagship 2 exclusively for galaxies labelled ``blue’’ (see Sect. \ref{subsec:simulations.flagship2}).
In the Flagship~2 catalogue, however, the fraction of galaxies classified as ``blue'' decreases with increasing redshift, resulting in a growing population of ``red'' and ``green'' satellites at higher redshifts.
These galaxies tend to be more strongly concentrated towards the halo centre than their blue counterparts, thereby enhancing the clustering signal at very small separations and introducing a systematic mismatch between the model and the data that becomes increasingly pronounced at high redshift.
Employing the \textit{profile rescaling} extension to improve the $\redchisq$ of the projected 2PCF fit, we obtain best-fit values of $c_\mathrm{norm} = 1.17$ and $R_\mathrm{max} = 1.89$ ($\redchisq = 1.77$), indicative of a satellite population exhibiting stronger small-scale clustering than would be expected if all galaxies in the Flagship~2 sample were classified as ``blue''.

\begin{figure*}[t!]
    \centering
    \includegraphics[width=0.8\textwidth]{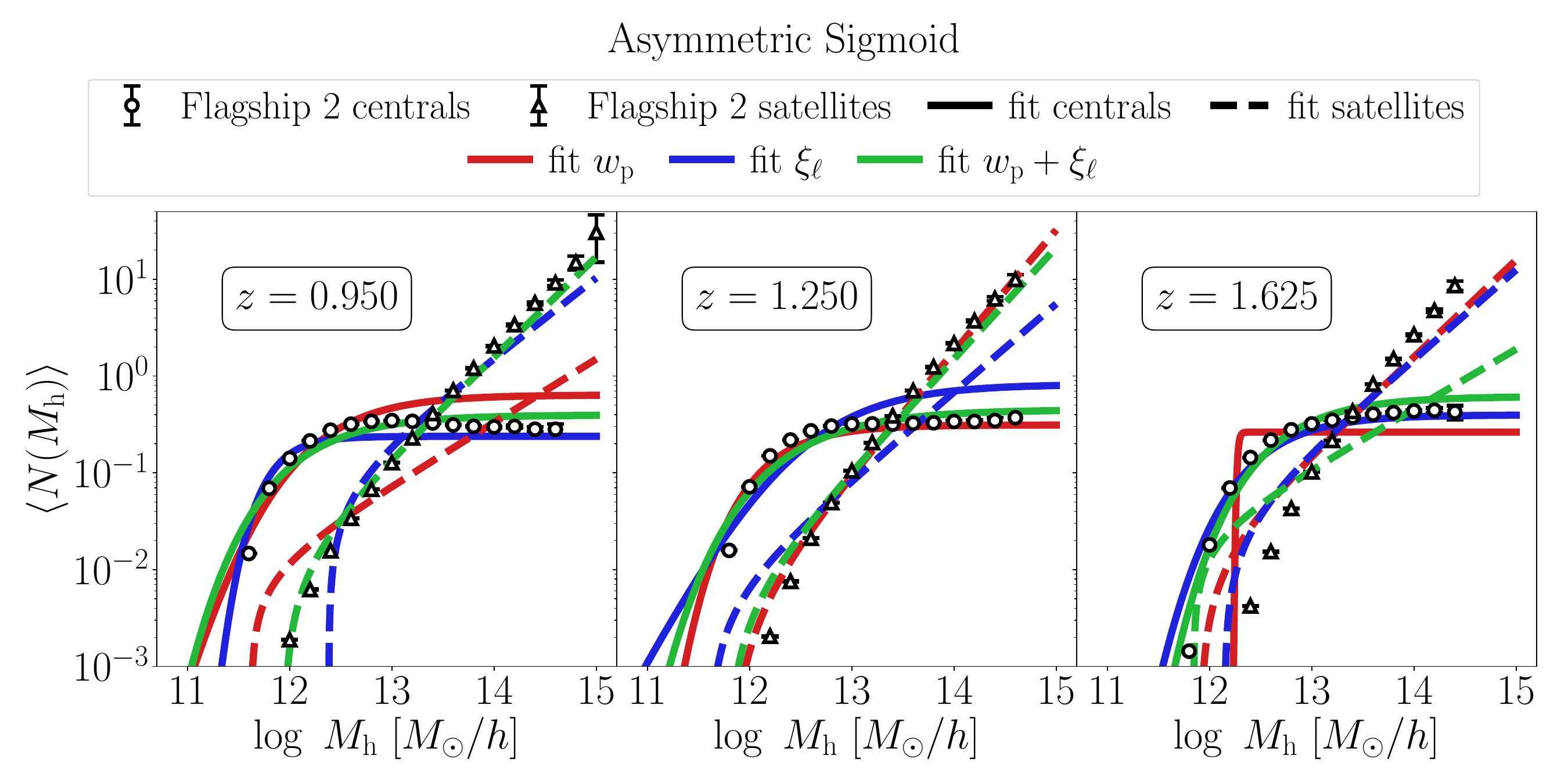}
    \caption{Best-fit HOD functions corresponding to the models shown in Fig.~\ref{fig:FS2_fit_redshifts}, for the Asymmetric Sigmoid parametrisation across three redshift bins ($z = 0.950$, $1.250$, and $1.625$). The mean halo occupation number $\langle N(M_\mathrm{h})\rangle$ is shown as a function of halo mass $\log M_\mathrm{h}\,[M_\odot/h]$, with circles and triangles representing the central and satellite contributions from the Flagship~2 simulation, respectively. Dark red, blue, and green lines correspond to the best-fit HODs obtained by fitting $w_\mathrm{p}$ only, $\xi_\ell$ only, and both statistics jointly, respectively, with solid and dashed lines indicating the central and satellite component, respectively.}
    \label{fig:FS2_HOD_redshifts}
\end{figure*}

In Fig.~\ref{fig:FS2_HOD_redshifts}, we show the resulting best-fit HOD shapes and compare them with a direct measurement of the HOD from Flagship 2.
A first notable observation is that best-fit models obtained from different observables do not necessarily converge to the same solution, a result that is not surprising in itself.
This simply reflects a certain degree of degeneracy between one-point statistics, such as the number density and the occupation functions, and two-point statistics such as $\wprp$ and $\xiell$.
Remarkably, the joint combination $\wprp+\xiell$ closely recovers the shape of the true occupation distribution of the fiducial model, with the exception of the satellite occupation in the highest-redshift bin.
These discrepancies, which in any case have little to no impact on the best-fit 2PCF, can in part be attributed to the colour-independent modelling of satellite positions in \scipiclite, which may not fully capture the complexity of the true satellite distribution.
These results further confirm that the joint analysis yields more robust, precise and accurate constraints than either observable considered individually.

\begin{figure}[t!]
    \centering
    \includegraphics[width=\columnwidth]{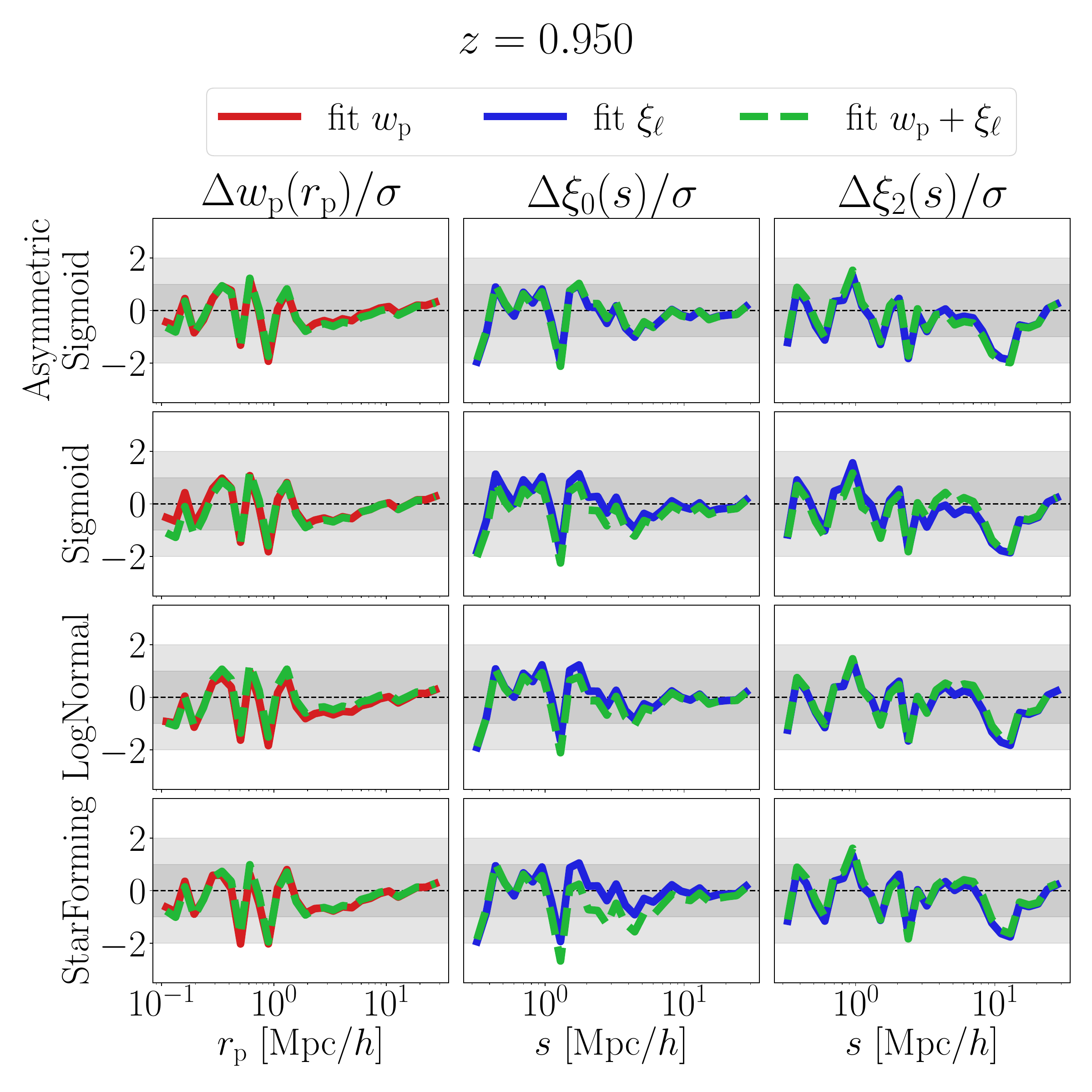}
    \caption{Residuals between the measured clustering of the Flagship 2 H$\alpha$ sample and the best-fit model at $z = 0.950$, normalised by the $1\sigma$ uncertainty, for different baseline HODs for central galaxies. From top to bottom, the rows show the Asymmetric Sigmoid, Sigmoid, LogNormal, and StarForming central occupation models. The columns show the projected correlation function $\wprp$ (left), the monopole $\xi_0(s)$ (middle), and the quadrupole $\xi_2(s)$ (right). Colours indicate the data used in the fit: $w_\mathrm{p}$ only (dark red), the multipoles $\xi_\ell$ only (blue), and $w_\mathrm{p} + \xi_\ell$ jointly (green).}
    \label{fig:FS2_fit_models}
\end{figure}

We then assess how robust the best fit is when varying the occupation function for centrals.
The panels of Fig.~\ref{fig:FS2_fit_models} show, for each HOD occupation function labelled on the left, the residuals between the data and the best-fit models.
The best-fit models are found to be remarkably consistent across all HOD parametrisations, demonstrating the robustness of the results with respect to the choice of HOD model.
This consistency is further reflected in the small differences in the $\redchisq$ values across models: the largest discrepancy arises for the StarForming parametrisation, whose \redchisq exceeds that of the Sigmoid (the lowest) by 0.08 when fitting $\wprp$ only.
This difference is statistically insignificant: for $\nu = 30 \ \mathrm{ (data \ points)} - 6 \ \mathrm{(free \ parameters)} = 24$ degrees of freedom, the \redchisq distribution has a standard deviation of $\sqrt{2/\nu} \approx 0.29$, meaning the discrepancy between the two models corresponds to only ${\sim}0.3\,\sigma$.
We note that, although some HOD models exhibit a decline in the central occupation at high halo masses, which makes the underlying distribution difficult to recover, the 2PCF is nonetheless reproduced equally well.
This is because the large-scale clustering amplitude is primarily set by the effective galaxy bias, which is in turn determined by the mean mass of the haloes hosting central galaxies: provided that this quantity is consistent across models, the clustering amplitude is recovered regardless of the detailed shape of the occupation function.
Different HOD prescriptions therefore yield statistically equivalent clustering, a degeneracy that complementary observables such as galaxy--galaxy lensing could help to break.

\subsection{HOD extensions}
\label{subsec:results.extensions}

\begin{figure}[t!]
    \centering
    \includegraphics[width=\columnwidth]{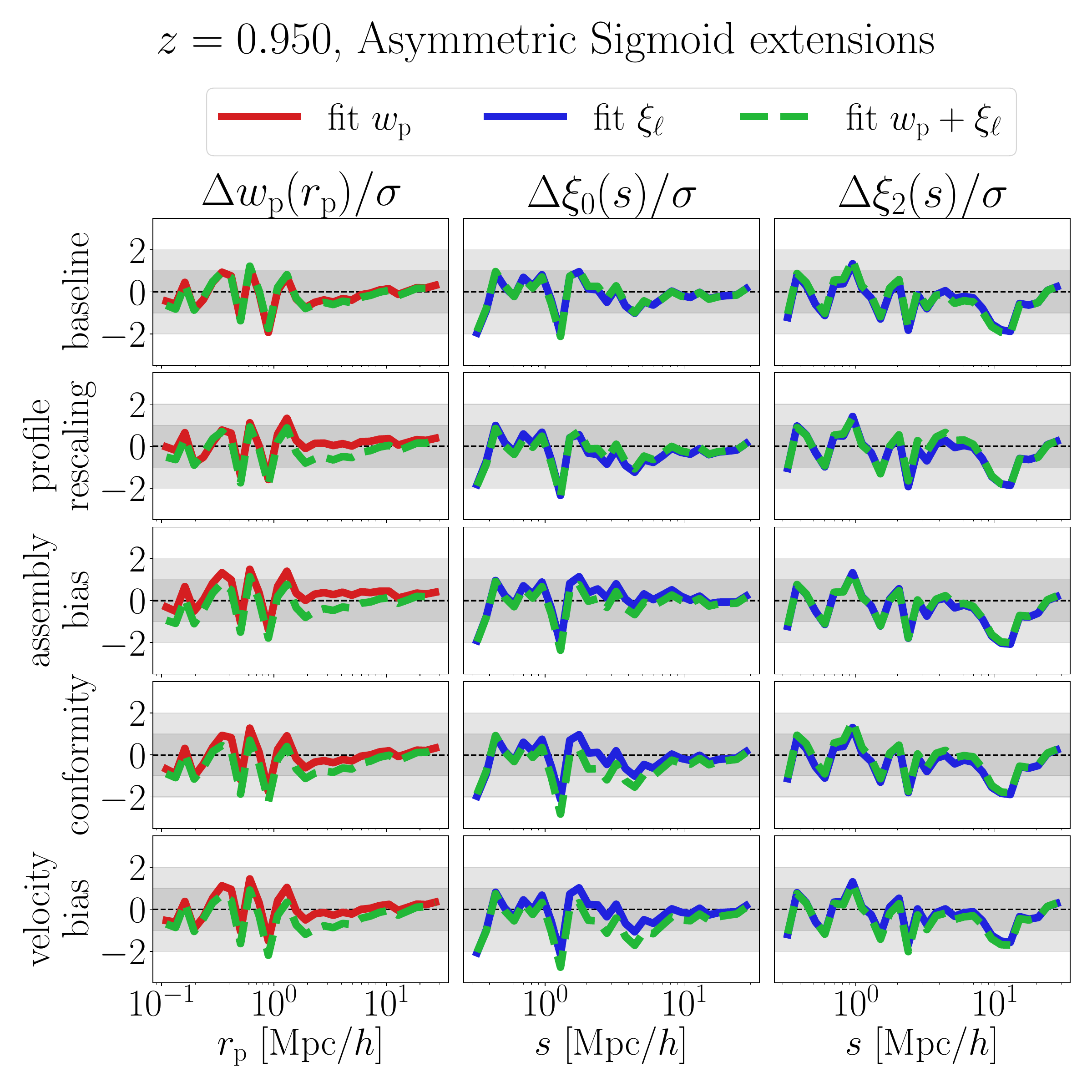}
    \caption{Same as Fig.~\ref{fig:FS2_fit_models}, but here we fix the central HOD to the Asymmetric Sigmoid and vary the HOD extension. From top to bottom, the rows show the baseline model and the models with profile rescaling, assembly bias, conformity, and velocity bias.}
    \label{fig:FS2_fit_extensions}
\end{figure}

We now turn our attention to the HOD extensions introduced at the beginning of this section.
The panels of Fig.~\ref{fig:FS2_fit_extensions} show a comparison of the residuals of the best-fit models obtained with the baseline Asymmetric Sigmoid HOD and its four extensions (profile rescaling, assembly bias, conformity, and velocity bias) at $z = 0.950$.

Once again, the best-fit models prove to be extremely robust against extensions of the standard HOD paradigm.
To quantify the differences between best-fit models more rigorously, we evaluate the statistical significance of the improvement in $\chi^2$ obtained when switching from the baseline model to each of its extensions.
At the redshift considered here, the largest discrepancy with respect to the baseline arises for the profile rescaling extension when fitting $\wprp$ alone (second row, left panel), yielding $\Delta\chi^2 = - 2.25$ at the cost of 2 additional free parameters, which is not statistically significant.

All extensions systematically lower the $\chi^2$ with respect to the baseline.
However, in most cases the improvement is not statistically significant once the additional free parameters are accounted for. 
Only two configurations stand out.
The first, already identified in Sect.~\ref{subsec:results.baseline}, is the fit to $\wprp$ at $z = 1.625$, where the profile rescaling extension reduces $\redchisq$ from 1.94 to 1.77, corresponding to $\Delta\chi^2 = -7.75$ for two additional parameters and a $p$-value of $0.021$ (${\sim}2.3\sigma$, assuming Wilks' theorem).
The second is the fit to $\xi_\ell$ at $z = 1.250$ with the LogNormal HOD, where the assembly bias extension yields $\Delta\chi^2 = -19.2$ for two additional parameters, corresponding to $p = 7\times10^{-5}$ (${\sim}4\sigma$).
This larger improvement is likely driven by the fact that the LogNormal HOD has fewer free parameters than other baseline models, so that the additional freedom introduced by the extension compensates for its more limited flexibility.
Even in these cases, however, the extended models still provide comparably poor fits to the data, with $\redchisq$ values well above unity.
This indicates that the residual discrepancies cannot be resolved by any of the extensions considered, and are more likely attributable to more fundamental limitations of the general modelling framework itself, as discussed in Sect.~\ref{subsec:results.baseline}.

\begin{figure*}[h!]
    \centering
    \includegraphics[width=\textwidth]{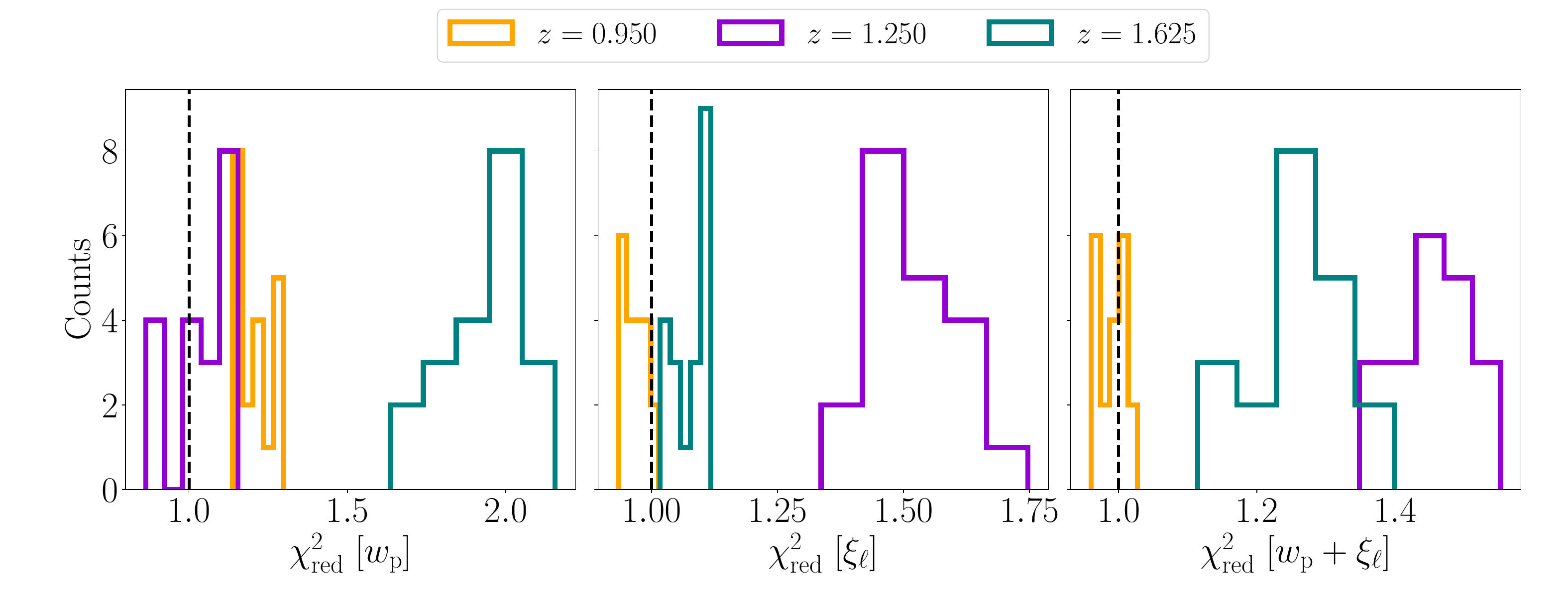}
    \caption{Distributions of the reduced chi-squared $\redchisq$ values obtained from fitting $w_\mathrm{p}$ only (left), $\xi_\ell$ only (centre), and both statistics jointly (right), across all HOD parametrisations and extensions considered in this work. Each histogram aggregates the $\redchisq$ values at redshifts $z = 0.950$ (orange), $z = 1.250$ (purple), and $z = 1.625$ (teal). The dashed vertical line marks $\redchisq = 1$ for reference.}
    \label{fig:chi2_plot}
\end{figure*}

This conclusion is supported by the $\redchisq$ distributions shown in Fig.~\ref{fig:chi2_plot}, where we show the $\redchisq$ values for all the fits we performed in this analysis.
Each histogram collects the $\redchisq$ values for the 4 HOD shapes used in this work combined with the baseline plus the 4 extensions (profile rescaling, assembly bias, velocity bias, conformity), for a total 20 points per redshift bin and observable.

The $\redchisq$ distributions at each redshift are quite narrow, with values clustered tightly around a common value regardless of the HOD parametrisation or extension employed.
This behaviour indicates that the goodness of fit is driven primarily by the data themselves, and by the scales and statistics being fitted, rather than by the specific choice of HOD model or the inclusion of additional extensions.
In other words, once the baseline modelling framework is fixed, the introduction of further free parameters provides only marginal improvement: if the baseline fit is poor, no extension is capable of capturing the residual discrepancies sufficiently to bring $\redchisq$ close to unity.
This is particularly evident for the cases reported above, where all models and extensions return $\redchisq$ values systematically above 1, pointing to a limitation that is inherent to the modelling framework rather than to the specific parametrisation adopted.

\section{AbacusSummit H$\alpha$ mock galaxy catalogues}
\label{sec:mocks}

Having established the robustness of our HOD fits to the 2PCFs of the Flagship~2 H$\alpha$ galaxy mock, we now turn to a complementary consistency test aimed at assessing whether the best-fit models, calibrated on small scales, are capable of reproducing the large-scale clustering signal as well.
This constitutes an important validation of the full pipeline: a model that accurately describes the 2PCF on quasi-linear and non-linear scales should, in principle, also provide a faithful representation of the power spectrum multipoles on larger scales, where linear theory and perturbative approaches are better understood.
Such a test is not only of theoretical interest, but is also of direct practical relevance for the generation of realistic galaxy mocks intended to mimic the clustering properties of galaxies observed in wide-field spectroscopic surveys.

To this end, we take the full set of best-fit HOD parameters for the joint $w_\mathrm{p}+\xiell$ fit obtained in the previous section, spanning all HOD parametrisations and extensions considered at a fixed redshift, and use them to populate the complete ensemble of the 25 Abacus realisations with Planck cosmology \citep{Abacus-Maksimova+21}.
For each realisation and each best-fit model, we generate a galaxy mock and measure the monopole, quadrupole, and hexadecapole of the power spectrum.

\begin{figure}[h!]
    \centering
    \includegraphics[width=0.9\columnwidth]{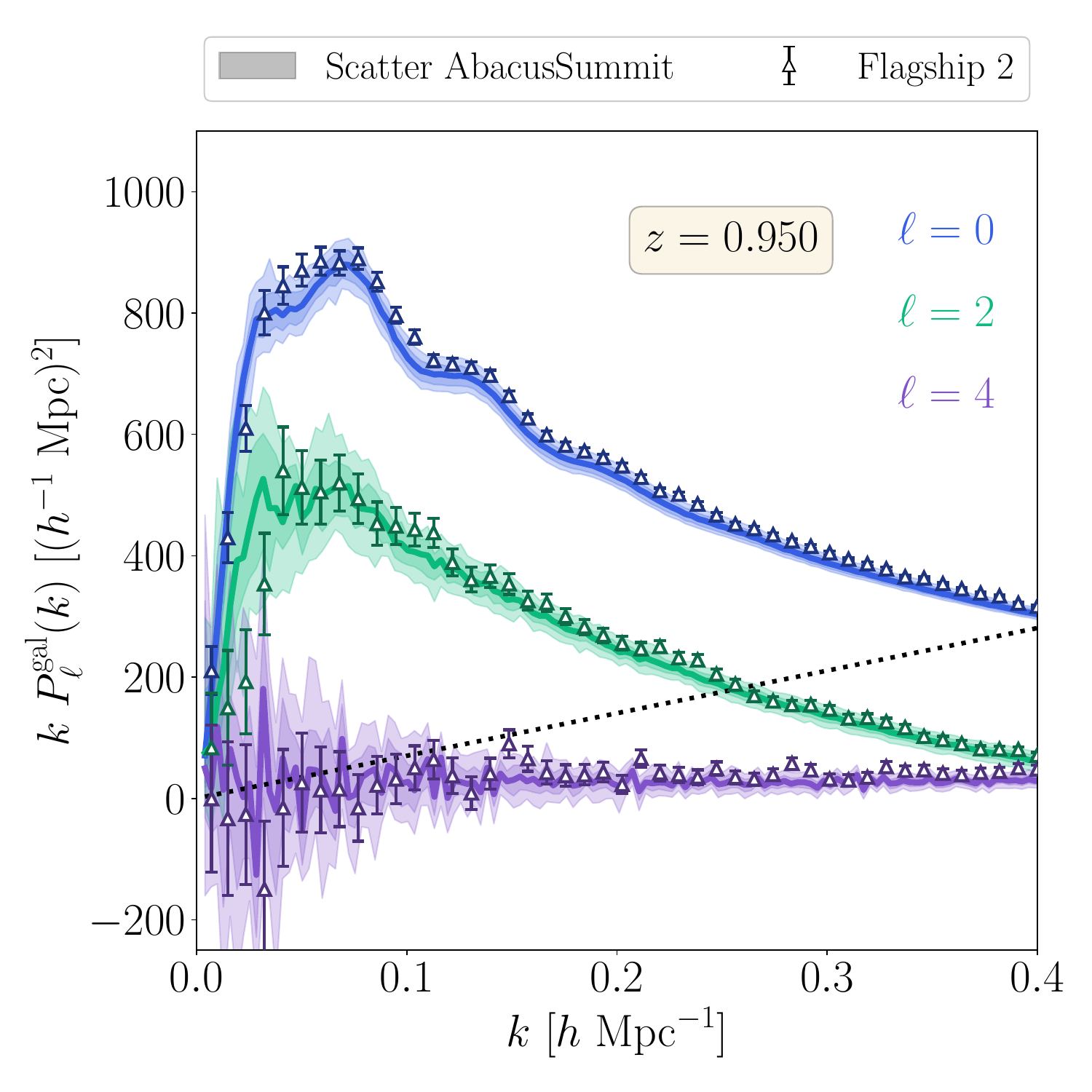}
    \caption{Power spectrum multipoles at $z = 0.950$ for all the best-fit HOD parametrisations (including extensions). Triangles show the monopole ($\ell = 0$, blue), quadrupole ($\ell = 2$, green), and hexadecapole ($\ell = 4$, purple) measured from the Flagship~2 galaxy mock, with error bars estimated under the Gaussian approximation for a volume corresponding to the full octant. Shaded bands show the $1\sigma$ and $2\sigma$ regions spanned by the ensemble of Abacus-based HOD models and extensions of the best-fit solutions from the joint $\wprp+\xiell$ of Flagship 2. The dotted black line indicates the shot noise contribution, subtracted from the monopole.}
    \label{fig:FS2_vs_Abacus_scatter}
\end{figure}

We present our results in Fig.~\ref{fig:FS2_vs_Abacus_scatter} at $z = 0.950$, focusing on the best-fit models obtained from the joint fit to $\wprp + \xiell$.
Triangles show the power spectrum multipoles measured from the Flagship~2 galaxy mock, with error bars estimated under the Gaussian approximation for a volume corresponding to the full octant \citep{Gaussian_covariance-Grieb+16}, and without accounting for the survey geometry.
Solid lines represent the mean monopole, quadrupole, and hexadecapole, computed over the full ensemble of HOD parametrisations and extensions, with darker and lighter shaded bands enclosing the $1\sigma$ and $2\sigma$ regions.
This ensemble comprises 500 mocks in total for each redshift, corresponding to 25 Abacus realisations combined across 4 HOD parametrisations and 4 extensions (plus the baseline).
The Poisson shot noise, shown as a dotted black line, has been subtracted from the monopole.

The Abacus predictions reproduce the Flagship~2 measurements well, although the monopole is systematically overestimated by ${\sim}1\sigma$.
A quantitative comparison is complicated by the window function affecting the Flagship~2 power spectrum, and the offset cannot be ascribed to a single modelling ingredient.
In addition, the Flagship~2 error bars are Gaussian and neglect the correlations between scales, which are significant and would reduce the significance of a coherent offset such as the one observed.
Furthermore, mocks generated from parameter sets sampled during the minimisation and lying within the $2\sigma$ region of the best fit (based on $\Delta\chi^2$) agree with Flagship~2 within $1\sigma$ (not shown).
This result is of particular importance for the subsequent steps of the analysis: the HOD models were calibrated on a small region of the Flagship~2 light cone, covering an area of $19.3\,\mathrm{deg}^2$, yet they are capable of accurately reproducing the clustering statistics at the largest scales when applied to a volume larger by a factor of up to 840 at the lowest redshift.
This remarkable consistency across scales and simulations provides strong validation of the fitting pipeline, and gives us confidence that the resulting Abacus-based galaxy mocks can be reliably employed for a broad range of downstream applications, including the testing and validation of theoretical models and analysis pipelines, and the propagation of HOD modelling uncertainties to wide-field analyses.

The full suite of Abacus galaxy mocks generated in this work, spanning all HOD parametrisations, extensions, and redshift bins considered, will be made publicly available on the CosmoHub platform \citep{CosmoHub} upon publication of this paper.
Additionally, the measurements of the power spectrum multipoles and 2PCFs presented here are available upon reasonable request to the authors.

\section{Conclusions}
\label{sec:conclusions}

In this work we have presented \scipiclite, a new Python library for populating dark matter haloes from cosmological simulations with galaxies following the HOD prescription.
\scipiclite inherits the computational efficiency of \textsc{SciPIC} while significantly extending its flexibility for galaxy clustering applications.
The library supports a wide variety of HOD parametrisations for both central and satellite galaxies, including a number of built-in models as well as the possibility of defining custom occupation functions, making it adaptable to different galaxy populations and scientific use cases.
Unlike simulation-specific frameworks such as \textsc{AbacusHOD}, \scipiclite is simulation-agnostic and supports light-cone outputs, ensuring portability across different \Nbody simulations.

Beyond the core occupation framework, \scipiclite offers a range of physically motivated modelling options, of which we summarise here those used in this work.
The built-in HOD models for central and satellite galaxies are fully customisable and support galactic conformity, while their parameters can be modulated on a halo-by-halo basis to naturally incorporate effects such as assembly bias.
Satellite positions are drawn from an NFW profile by default, with custom density profiles available to better reproduce hydrodynamical simulations.
Satellite velocities are then obtained by solving the Jeans equation consistently with the chosen profile, with full control over the velocity bias parameters.
The library includes tools for measuring 2PCFs, power spectrum multipoles and density fields, making it a self-contained environment for end-to-end galaxy clustering analyses.
The code also provides minimisation pipelines based on the differential evolution algorithm \citep{Differential_evolution-Storn-Price+97}, which can be used to fit HOD parameters to target clustering measurements.

We tested and validated \scipiclite by fitting the small-scale 2PCFs of H$\alpha$-selected galaxies from the Flagship~2 simulation \citep{Flagship-Castander+24} across three redshift bins, from $z=0.85$ to $z=1.80$.
We modelled this signal by populating the Abacus halo catalogues \citep{Abacus-Maksimova+21} with mock galaxies.
We then used the differential evolution pipeline contained in \scipiclite to find the parameters that minimise the $\chi^2$ against the Flagship~2 measurements.
We performed the fit for three observables: $w_\mathrm{p}(r_\mathrm{p})$, $\xi_\ell(s)$, and their combined analysis.
In all cases, we explicitly accounted for the cross-covariance between statistics.

We presented results for a variety of central galaxy HOD parametrisations, as well as for several extensions to the standard HOD paradigm, including assembly bias, velocity bias, conformity, and profile rescaling.
The best-fit 2PCF models were found to be remarkably robust across the different HOD parametrisations considered, with the goodness of fit being driven primarily by the data and the scales being modelled rather than by the specific choice of HOD model or the inclusion of additional free parameters.
This is reflected in the $\redchisq$ values, which cluster tightly around a common value for each redshift bin and fitting configuration, regardless of the parametrisation adopted.
This reveals degeneracies in the HOD that two-point clustering cannot break; galaxy--galaxy lensing, already supported by \scipiclite, could help.
While some HOD extensions were found to yield statistically significant improvements over the baseline models in specific cases, these improvements were insufficient to fully resolve the residual discrepancy between the obtained $\redchisq$ and the ideal expectation of $\redchisq = 1$.
This indicates that the residual limitations of the fits are unlikely to arise from the choice of HOD parametrisation, but are instead attributable to more fundamental aspects of the modelling framework that fall outside the scope of the extensions considered in this work.

We exploited the best-fit HOD parameters obtained from our minimisation pipeline to generate a new suite of full-box galaxy mocks within the Abacus simulation framework, comprising a total of 1\,500 catalogues spanning 3 redshift snapshots, 4 HOD parametrisations, 4 extensions (plus baseline), and 25 independent realisations.
The power spectrum multipoles measured from this mock suite were found to be in good agreement with those of the Flagship~2 galaxy mock across all scales probed, including the large-scale regime well beyond the range of scales used in the fitting procedure.
This consistency between small-scale fits and large-scale predictions provides a stringent validation of the \scipiclite library and the newly developed pipeline.


\begin{acknowledgements}  
    The authors thank Marcos Pellejero, Jonas Chaves-Montero and Sladjana Radinovic for useful discussions. This work made use of the computing infrastructure, data storage and technical support provided by the Port d’Informació Científica (PIC), Barcelona, Spain. GP, MC and GG acknowledge support from the Spanish Ministerio de Ciencia, Innovación y Universidades, project PID2021-128989NB. Our research made use of \texttt{matplotlib}, a \texttt{Python} library for publication-quality graphics \citep{Hun2007}.
\end{acknowledgements}

\bibliographystyle{aa}
\bibliography{bibliography}

\begin{appendix}

\section{\scipiclite: advanced features}
\label{app:scipiclite.features}

While the core functionalities of \scipiclite have already been described in Sect.~\ref{sec:scipiclite}, here we provide an overview of more advanced features that extend the capabilities of the library, significantly broadening its applicability:

\begin{itemize}
    \item \textbf{Non-Poisson satellite statistics}: support for super-Poisson (negative binomial) and sub-Poisson \citep[binomial, rounded to the nearest inverse integer,][]{HOD_nonPoisson-Vos-Gines+24} satellite occupation distributions, allowing for a more faithful representation of the stochasticity of satellite counts within haloes;
    
    \item \textbf{Redshift evolution}: HOD parameters can evolve with redshift on a halo-by-halo basis, with dedicated functions to predict the expected redshift dependence for flux-limited samples;
    
    \item \textbf{Redshift errors}: errors of arbitrary functional form can be applied to galaxy redshifts through a flexible \texttt{SciPy}-based interface, with full user control over the error model and its parameters;
    
    \item \textbf{Abundance matching}: functions to implement luminosity functions and stellar-to-halo mass relations, including the mass-matching procedure adopted in this work for the population of Abacus haloes;
    
    \item \textbf{Cosmological calculations}: a suite of utilities for basic cosmological quantities, including comoving distances, growth factors, Eisenstein \& Hu linear power spectra \citep{EisensteinHu_1998}, densities, virial radii, and mass functions;

    \item \textbf{Field measurements}: routines for the measurement of density and tidal fields, and halo-by-halo large-scale bias \citep{halo_by_halo_bias-Paranjape+17};

    \item \textbf{Survey footprint}: interface with \texttt{healpy} to generate random catalogues for generic footprints;

    \item \textbf{Statistics}: interface with \texttt{Corrfunc} \citep{Corrfunc} to measure 2PCFs, cross-2PCFs, excess surface densities, and built-in routines to measure power spectra, HODs and mass functions;

    \item \textbf{Utilities}: interface with \texttt{pyfftlog} to compute Fourier transforms and switch from power spectra to 2PCFs; analytical halo model predictions.
\end{itemize}


\section{Iterative MCMC fit}
\label{app:scipiclite.iterative_fit}

In this appendix, we return to the iterative MCMC fitting method introduced at the end of Sect.~\ref{subsec:scipiclite.pipeline}.
\cite{Iterative_emulator-Pellejero+19} showed that this approach recovers the posterior contours accurately while reducing the number of likelihood evaluations by up to a factor of $\sim 100$.
As a demonstration of its applicability in our context, we apply it to a single case: the fiducial Asymmetric Sigmoid HOD (Eq.~\ref{eq:scipiclite.centrals.asymmetricsigmoid}) at $z=0.950$.
Given the limited footprint of the sample (19.3 deg$^2$), we reduce the dimensionality of the parameter space by fixing $\logMcut=12$ and $\log\nu=1$.
Since these parameters are expected to be only weakly constrained by the data, leaving them free would expose the posterior to prior volume projection effects.
The procedure begins with the minimisation of the $\chi^2$ (Eq.~\ref{eq:scipiclite.pipeline.chi2}) via differential evolution, during which every sampled point in parameter space is stored together with its likelihood value.
The 50 lowest-$\chi^2$ points and 50 additional random points are extracted from this sample and 
used to build a surrogate model of the posterior using a Gaussian process emulator, as implemented in \texttt{scikit-learn}\footnote{\url{https://scikit-learn.org/}}.
An MCMC is subsequently run on this surrogate with \texttt{emcee}\footnote{\url{https://emcee.readthedocs.io/en/stable/}}, and the chain is considered converged once its length exceeds 50 times the autocorrelation time of each parameter.
Two new points are then drawn at random from the converged chain, their exact likelihood is computed, and they are added to the training set.
This cycle of emulator training, chain sampling and training-set augmentation is iterated until the Kullback--Leibler divergence \citep{KL} between consecutive posteriors remains below 0.1 for 10 consecutive iterations.

\begin{figure}[h!]
    \centering
    \includegraphics[width=\columnwidth]{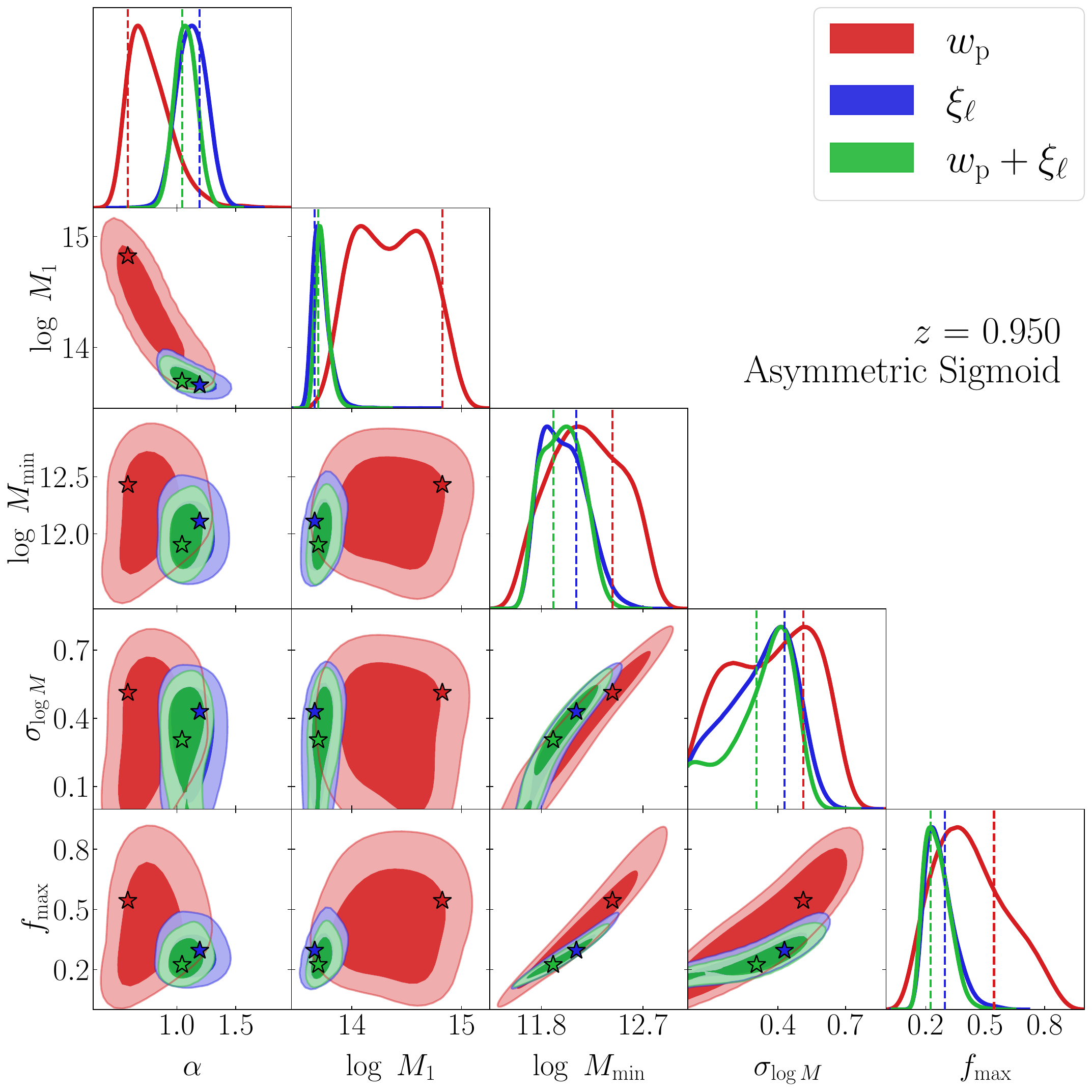}
    \caption{Corner plot showing the posterior distributions of the HOD parameters for the Asymmetric Sigmoid central galaxy parametrisation at $z = 0.950$, obtained from the iterative fit procedure described in Sect. \ref{subsec:scipiclite.pipeline}. Dark red, blue, and green contours correspond to models obtained by fitting $\wprp$ only, $\xiell$ only, and both statistics jointly, respectively. Coloured stars represent the best-fit values obtained by the previous $\chi^2$ minimisation with the differential evolution algorithm.}
    \label{fig:FS2_fit_iterative}
\end{figure}

In Fig.~\ref{fig:FS2_fit_iterative} we show the resulting contour plots for three different probes, accompanied by their best-fit values estimated with the differential evolution minimiser approach.
Notably, all three contours were obtained from fewer than 300 direct likelihood evaluations, at a substantially lower computational cost than a standard MCMC analysis would require.
Strong degeneracies are found that are responsible for some residual prior volume projection effects, as the best-fit values do not always lie within the marginalised $1\sigma$ contours.
We verified this by comparing the maximum a posteriori point of the MCMC chains with the best-fit parameters obtained from the minimiser, finding excellent agreement.
The effect is strongest for the $\wprp$-only fit, where $\logMone$ and $\alpha$ are poorly constrained and almost limited by the priors; the $\xiell$ and joint best fits agree well with the marginalised posteriors.
The parameter contours are visibly tighter for the joint fit of $w_\mathrm{p}(r_\mathrm{p})$ and $\xi_\ell(s)$ compared to either statistic alone, consistent with the larger amount of information encoded in their combination.
The parameters which are most constrained are $\logMmin$ and $\logMone$ which together are responsible for the clustering amplitude: in particular, the former is responsible for the large-scale bias while the ratio of the latter to the former has an analogous impact at small scales, as it fixes the total satellite fraction.
Central HOD parameters are also found to be tightly correlated:
$\logMmin$ and $\sigmalogM$ jointly set the mean mass of haloes hosting central galaxies: higher values of $\sigmalogM$ and $\logMmin$ would result in very similar clustering amplitude, therefore explaining the positive correlation between the two parameters.
Finally, $f_\mathrm{max}$ exhibits a degeneracy with $\log M_\mathrm{min}$, which however is not due to clustering. $f_\mathrm{max}$ is a derived parameter fixed by the requirement of matching the observed number density, and $\log M_\mathrm{min}$ is the other free parameter mainly governing that quantity. The correlation between the two reflects the constraint imposed by the number density rather than a genuine physical degeneracy in the clustering signal.

\end{appendix}

\end{document}